\documentclass[superscriptaddress,twocolumn,preprintnumbers]{revtex4}
\usepackage[T1]{fontenc}
\usepackage[latin1]{inputenc}
\usepackage{amsmath}
\usepackage{amssymb}
\usepackage{dsfont}
\usepackage{color}
\makeatletter
\usepackage{amssymb}
\usepackage{commath}
\usepackage{graphicx}
\usepackage{epstopdf}
\usepackage{epsfig}
\usepackage{subfigure}
\usepackage{relsize}
\usepackage[colorlinks,urlcolor=blue,citecolor=blue]{hyperref}
\makeatother

\usepackage{cancel,ifthen}
\usepackage[normalem]{ulem}

\newcommand{\me}[1]{{\it #1}}

\newcommand{\comment}[2][NoInPuT]{\ifthenelse{\equal{#1}{NoInPuT}}{}{{\color{blue}\sout{#1}}}{\color{red} #2}}
\begin{document}

\title{Synthetic superfluid chemistry with vortex-trapped quantum impurities}

\author{Matthew Edmonds}
\affiliation{Department of Physics \& Research and Education Center for Natural Sciences, Keio University, Hiyoshi 4-1-1, Yokohama, Kanagawa 223-8521, Japan}
\author{Minoru Eto}
\affiliation{Department of Physics, Yamagata University, Kojirakawa-machi 1-4-12, Yamagata, Yamagata 990-8560, Japan}
\author{Muneto Nitta}
\affiliation{Department of Physics \& Research and Education Center for Natural Sciences, Keio University, Hiyoshi 4-1-1, Yokohama, Kanagawa 223-8521, Japan}

\date{\today{}}

\begin{abstract}\noindent
We explore the effect of using two-dimensional matter-wave vortices to confine an ensemble of bosonic quantum impurities. This is modelled theoretically using a mass-imbalanced homogeneous two component Gross-Pitaevskii equation where each component has independent atom numbers and equal atomic masses. By changing the mass imbalance of our system we find the shape of the vortices are deformed even at modest imbalances, leading to barrel shaped vortices; which we quantify using a multi-component variational approach. The energy of impurity carrying vortex pairs are computed, revealing a mass-dependent energy splitting. We then compute the excited states of the impurity, which we in turn use to construct `covalent bonds' for vortex pairs. Our work opens a new route to simulating synthetic chemical reactions with superfluid systems.
\end{abstract}

\preprint{YGHP-20-07}

\maketitle

\section{Introduction}
Quantized vortices represent the fundamental excitations of superfluids -- atomic gases formed from interacting particles that exhibit non-viscous transport phenomena. Early experimental work revealed the nature of superfluid vortices \cite{chevy_2000,raman_2001,abo_2001} in these systems, which has also spurred theoretical interest, since the Gross-Pitaevskii formalism facilitates the accurate modelling of the phenomenology of superfluid alkali gases \cite{fetter_2009,dalfovo_1999}. 

Atomic Bose-Einstein condensates represent exceptionally pure physical systems, since it is possible to experimentally realize ground states with almost no non-condensate atoms present. This gives a unique opportunity to study the role of impurities in these systems, with recent pioneering experiments realizing both bosonic \cite{jorgensen_2016,guang_2016} and fermionic \cite{schirotzek_2009} polarons. Complementary to this, experiments with binary condensates have now achieved the trapping of one matter-wave inside another, here a degenerate Fermi gas formed of $^6$Li atoms was confined inside a Bose-Einstein condensate of $^{133}$Cs atoms \cite{desalvo_2017}. Attractive atomic interactions provide another opportunity to understand the interaction of impurities with solitary waves in the nonlinear regime \cite{kalas_2006,sacha_2006,edmonds_2019}.

Homogeneous systems represent an important testing ground for theoretical ideas, and have been under active experimental pursuit. Experiments with cold atomic gases typically use magnetic or optical traps to provide spatial confinement. However, recently it has become possible to confine atomic gases inside potentials that realize an effective box trap, leading to homogeneous bosonic ground states in one \cite{meyrath_2005,es_2010}, two \cite{jalm_2019}, and three-dimensional systems \cite{gaunt_2013,navon_2016}. Degeneracy has also been achieved with Fermi gases confined in homogeneous potentials \cite{hueck_2018}. Homogeneous systems offer a unique opportunity to investigate superfluidity in a clean and precise way \cite{stagg_2014} and understand the non-equilibrium physics of uniform systems \cite{chomaz_2015}.

Quantum mechanical gases manifest superfluidity by the nucleation of quantized vortices when the gas undergoes rotation. Typically this leads to the formation of the {\it Abrikosov} lattice at equilibrium, however recent work has revealed that homogeneous \cite{adhikari_2019}, multi-component \cite{kasamatsu_2003,mingarelli_2019} and density-dependent \cite{edmonds_2020} gauge theories all exhibit novel vortex configurations. Since individual vortices are topologically protected, they represent an intriguing candidate for hosting impurities within the matter-wave \cite{braz_2020}. In the context of the strongly interacting Helium fluids, charged impurities in the form of electrons were originally proposed as an extension to the Gross-Pitaevskii formalism \cite{clark_1965}, which has recently found renewed interest due to the ability to accurately numerically simulate the Gross-Clark model \cite{villois_2018}.

The weakly correlated regime represented by atomic Bose-Einstein condensates provides an important testing ground for understanding the dynamics of superfluid vortices \cite{barenghi_2016}, which in turn provides new theoretical \cite{tsatsos_2016,white_2014} and experimental \cite{kwon_2016} insight into quantum turbulence and non-equilibrium phenomenology \cite{liu_2018}. Understanding the complex dynamics of superfluids is assisted by knowledge of the few-body dynamics of vortices, in particular the realization of vortex dipoles \cite{neely_2010} continues to provide important information \cite{cawte_2019,yang_2019}, as well as the observation of solitonic vortices \cite{donadello_2014} in elongated superfluids. Multiple vortex states are generated by rotating the atomic cloud \cite{castin_1999,tsubota_2002,cozzini_2003}, leading at large rotation to the formation of the Abrikosov vortex lattice \cite{oriordan_2016,oriordan_2016a}.
\begin{figure*}[t]
\includegraphics[width=0.9\textwidth]{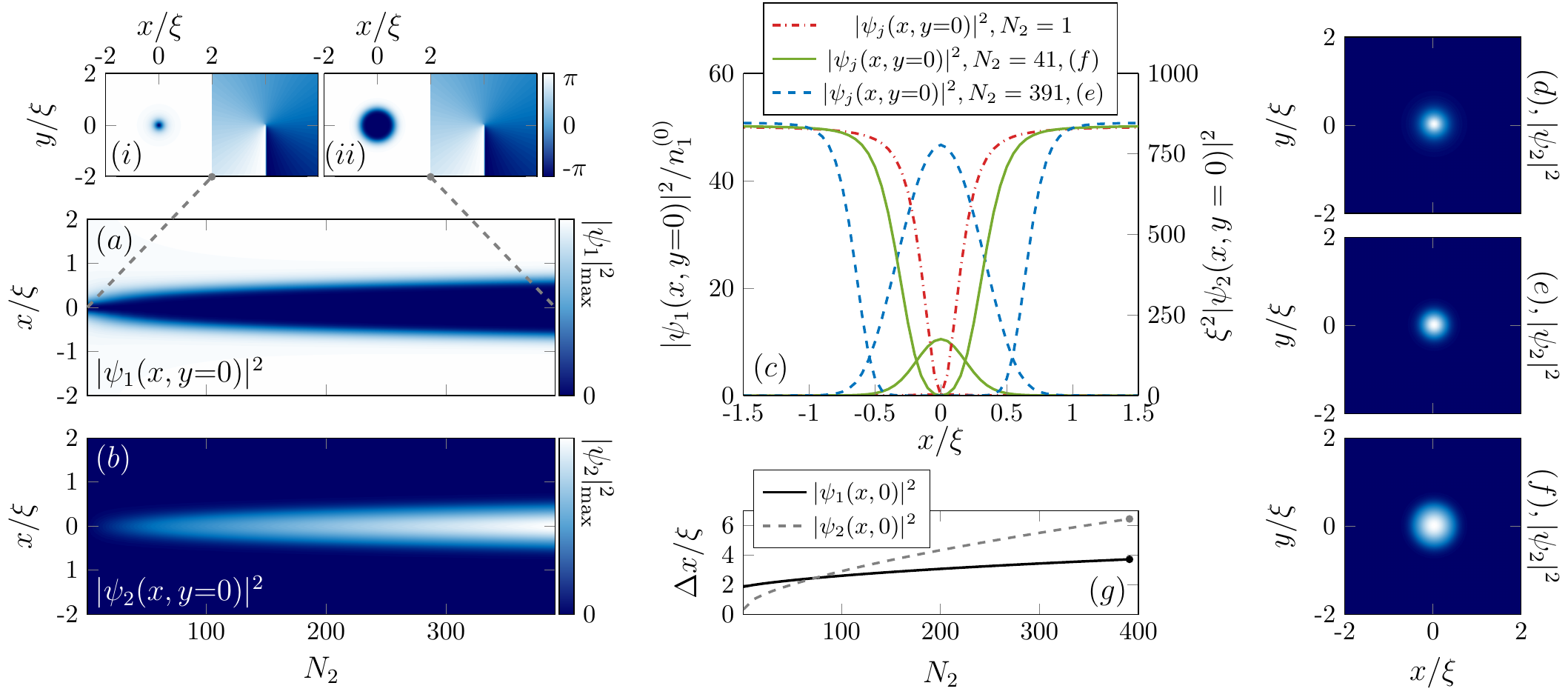}
\caption{\label{fig:single}(color online) Single vortex ground states. Panels (a) and (b) show ground state cross-sections $|\psi_j(x,y=0)|^2$ of the superfluid and impurity as a function of $N_2$ respectively. Corresponding density $|\psi_1|^2$ and phase $\vartheta_1=\tan^{-1}(\text{Im}(\psi_1)/\text{Re}(\psi_1))$ distributions are displayed for $N_2=1\text{ and }391$ in panels (i,ii). (c) shows density cross-section for each component for different numbers of impurities, while panels (d-f) show impurity densities $|\psi_2|^2$ for $N_2=\{1,41,391\}$ respectively. Panel (g) shows the size (standard deviation) information of the superfluid-impurity system. Throughout we have taken $N_1=4{\times}10^3$.}
\end{figure*}

Quantum mechanical gases with several coupled interacting internal degrees of freedom present an ongoing opportunity to gain a deeper insight into exotic forms of superfluidity in microscopic systems \cite{lewenstein_book}. 
Condensates with coherent couplings have also been an ongoing focus, with the dynamics of vortices in these models showing phenomenology analogous to high energy physics \cite{Son:2001td,Kasamatsu:2004tvg,Eto:2012rc,Eto:2013spa,Cipriani:2013nya,Tylutki:2016mgy,eto_2018,eto_2020,Kobayashi:2018ezm}. Condensates with spin degrees of freedom have been shown to possess rich vortex physics, including stable multiply charged vortices \cite{leanhardt_2002,mizushima_2002} in the ferromagnetic phase of these systems, attributed to the existence of both mass and spin current, leading to two individual forms of circulation. It has also been possible to consider ``half'' quantized vortices in the antiferromagnetic phase, where the circulation is still quantized but the unit of quantization becomes a fraction leading again to unique lattice phases \cite{martikainen_2002,mukerjee_2006}, as well as exotic vortex structures with nontrivial core structures which have been studied in spinor condensates \cite{leonhardt_2000,ohmi_1998,kobayashi_2012,lovegrove_2014,borgh_2016,kawaguchi_2012}. In multi-component systems, vortices are usually core-less \cite{Kasamatsu2005} -- the cores of vortices in one component are filled by the other component(s). This leads to non-trivial vortex-vortex interactions \cite{Eto:2011wp,Kasamatsu:2015cia} and consequently rich phase structures of vortex lattices other than the Abrikosov vortex lattice \cite{Mueller2002,Kasamatsu2003,Kasamatsu2005,Cipriani:2013nya,Yang:2020wes}.

Multi-component systems can also host more elaborate topological structures such as monopoles \cite{stoof_2001,ruostekoski_2003} consisting of a radially varying spin structure, which have been realized in the ferromagnetic spin-1 phase of a $^{87}$Rb condensate \cite{ray_2014}. It is also possible to consider even more elaborate and exotic excitations, such as skyrmions in 2D \cite{choi_2012} and 3D \cite{Khawaja:2001zz,Ruostekoski:2001fc,Battye:2001ec,Nitta:2012hy,kawakami_2012} which arise as a spatially varying spin deformation of the ground state of spinor systems as well as knots that are characterized in terms of the Hopf invariant, a topological charge unique to these excitations \cite{kawaguchi_2008,hall_2016,ollikainen_2019}. Multi-component systems also host exotic composite topological excitations such as D-brane solitons -- vortices ending on a domain wall \cite{Kasamatsu:2010aq,Kasamatsu:2013lda,Kasamatsu:2013qia}. Thus, multi-component superfluids provide an ongoing resource for topology, facilitating the creation of these exotic excitations in a stable and controlled environment. 

We consider a two-component system in which only the first component is condensed, while the second feels the superfluid component as an effective potential. We study how impurity atoms can be trapped and manipulated by individual and pairs of vortices in homogeneous superfluids. We specifically consider the situation where each component can have different atom numbers, but the atomic mass of both components are equal. It is also possible to generalize this model to consider noninteracting fermionic impurities, which entails having a set of wave-like equations coupled to a superfluid component. Such a situation was proposed previously by Ref. \cite{rutherford_2011} who modelled the experiment of Palzer et al., \cite{palzer_2009} who studied the transport properties of spin carrying impurity atoms through a one-dimensional Bose gas. By changing the number of atoms in the individual components, we reveal the unusual phenomenology of this system; including distorted vortex profiles and mass-dependent splitting of the impurities energy. We then explore the properties of the excited states of the impurity, including the possibility of synthetic chemical bonds. 

The paper is organized as follows. In Sec.~\ref{sec:tm} we introduce the model that describes our binary system, before examining the physics of individual vortices coupled to differing numbers of impurities in Sec.~\ref{sec:nr}, including a variational description of the system to explain the effect of the impurities on the vortex. We then compute the energy of pairs of vortices carrying impurities with different combinations of charges. Section.~\ref{sec:cb} examines the exited states of the vortex-impurity system, revealing the possibility of vortex chemistry with superfluids. In Sec.~\ref{sec:cc} we summarize our findings.                    
   
\section{\label{sec:tm}Theoretical Model}
We consider a system of $N_1$ atoms forming a Bose-Einstein condensate described by the state $\psi_1$, self-consistently collisionally coupled to a second component of $N_2$ impurity atoms, described respectively by the state $\psi_2$. We consider the case where both components have equal masses $M$. Then, after projecting the three-dimensional binary system into two-dimensions by integrating out the axial degree of freedom one obtains the coupled Gross-Pitaeveskii equations  
\begin{subequations}\label{eqn:gpe}
\begin{align}
i\hbar\frac{\partial\psi_1}{\partial t}&=\bigg[-\frac{\hbar^2}{2M}\nabla^2+g_{\rm 11}|\psi_1|^2+g_{\rm 12}|\psi_2|^2\bigg]\psi_1,\\
i\hbar\frac{\partial\psi_2}{\partial t}&=\bigg[-\frac{\hbar^2}{2M}\nabla^2+g_{\rm 12}|\psi_1|^2\bigg]\psi_2,
\end{align} 
\end{subequations} 
where $g_{\rm 11}=\hbar^2 a_s/Ma_{\rho}^2$ encapsulates the scaled three-dimensional $s$-wave scattering length $a_s$ of the atoms with mass $M$, while $g_{\rm 12}$ defines the strength of the collisional coupling between the superfluid and the impurity. The normalization of each component obeys $\int d{\bf r}|\psi_j({\bf r})|^2=N_j$, while the mass imbalance is $N_2/N_1$. Meanwhile, the total (conserved) energy associated with the Gross-Pitaevskii model represented by Eqs.~\eqref{eqn:gpe} is defined as
\begin{equation}\label{eqn:en}
E{=}\int d{\bf r}\bigg[\frac{\hbar^2}{2M}\sum_j|\nabla\psi_j|^2{+}\frac{g_{\rm 11}}{2}|\psi_{\rm 11}|^4{+}g_{\rm 12}|\psi_1|^2|\psi_2|^2\bigg].
\end{equation}
A single isolated two-dimensional vortex constitutes a hole on an otherwise homogeneous background, so it is important to separate the energy associated with the background and that of the vortex itself. As such we can write the total density of the superfluid as \cite{pethick_2002} $|\psi_1|^2=n_{1}^{(0)} - (n_{1}^{(0)} - |\psi_1|^2)$ where the background is $n_{1}^{(0)}$ (i.e. the value of the superfluid density far from the vortex core) while the term in brackets accounts for the vortex core region. Then, the atom number for the superfluid component can be expressed as $N_1=\mathcal{A}_{\rm 2D}n_{1}^{(0)}{-}\int d{\bf r}\big(n_{1}^{(0)}{-}|\psi_1|^2\big)$, where $\mathcal{A}_{\rm 2D}$ defines the area of the first component. The atom number $N_1$  is comprised of a constant background atom number $\mathcal{A}_{\rm 2D}n_{1}^{(0)}$ and the atom number of the vortex core. To obtain the energy of a single vortex in the two-component system, the energy associated with the homogeneous (vortex free) state is $E_0/n_{1}^{(0)}{=}g_{\rm 11}\mathcal{A}_{\rm 2D}n_{1}^{(0)}/2 - \int d{\bf r}\big[2(n_{1}^{(0)}{-}|\psi_1|^2){-}g_{\rm 12}|\psi_2|^2\big]$ and by writing $E_0/n_{1}^{(0)}$ higher order terms have been dropped, assuming that the size of the system $L_{x,y}$ satisfies $L_{x,y}\gg\xi$, where $\xi=\hbar/\sqrt{M\mu_1}$ is the size (healing length) of the vortex core. Then, the energy associated with the vortex $E_{\rm v+i}=E-E_{0}$ can be found using our expression for $E_0$ and Eq.~\eqref{eqn:en}, giving $E_{\rm v+i}=\sum_{j=1}^{2}E_{j}^{\rm kin}+E_{\rm 11}^{\rm vdW}+E_{\rm 12}^{\rm vdW}$ where the individual contributions can be written as
\begin{subequations}
\begin{align}
\label{eqn:ekin}E_{j}^{\rm kin}&=\frac{\hbar^2}{2M}\int d{\bf r}|\nabla\psi_j|^2,\\
\label{eqn:vdw1}E_{\rm 11}^{\rm vdW}&=\frac{g_{\rm 11}}{2}\int d{\bf r}\big(|\psi_1|^2 - n_{1}^{(0)}\big)^2,\\
\label{eqn:vdw2}E_{\rm 12}^{\rm vdW}&=g_{\rm 12}\int d{\bf r}\big(|\psi_1|^2 - n_{1}^{(0)}\big)|\psi_2|^2.
\end{align} 
\end{subequations}
Equations \eqref{eqn:ekin}-\eqref{eqn:vdw2} will be used in Sec.~\ref{sec:nr} to calculate the energy of different vortex configurations.
\begin{figure}[t]
\includegraphics[width=\columnwidth]{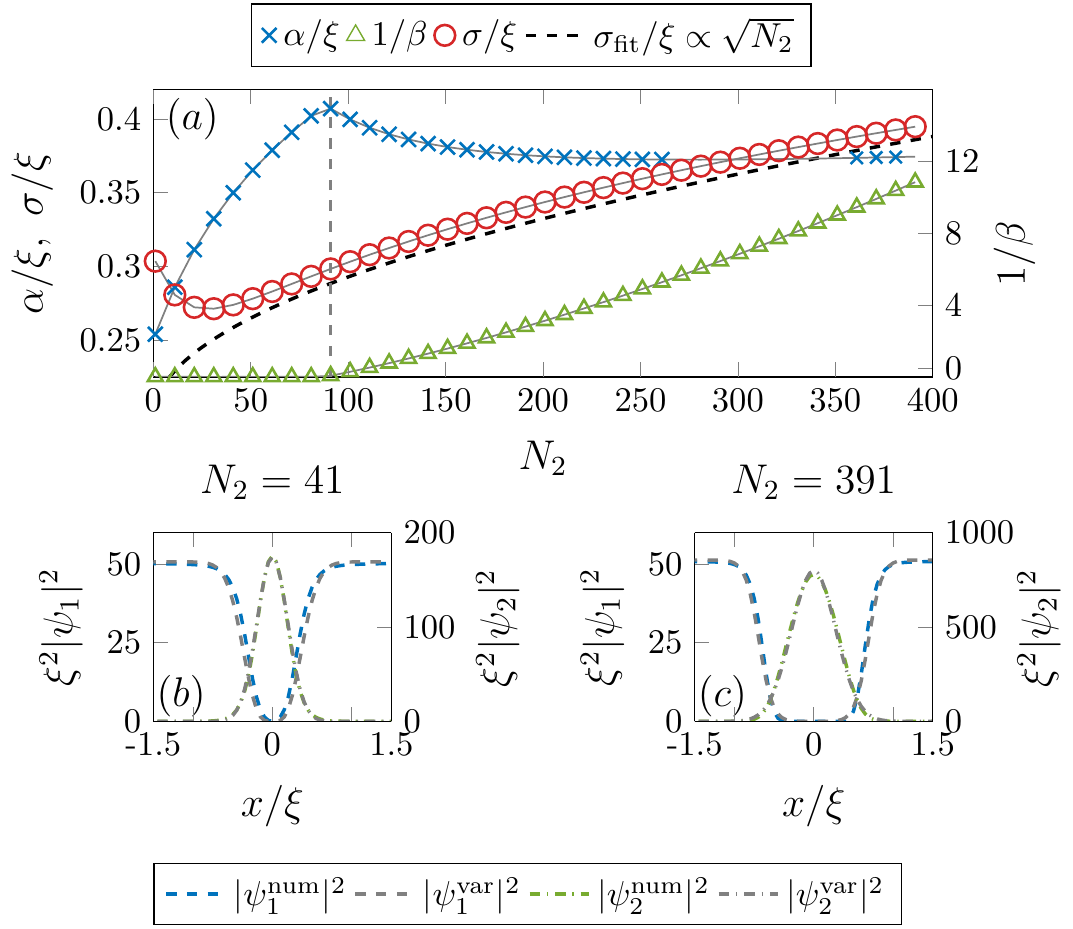}
\caption{\label{fig:fit}(color online) Variational parameters. The variational parameters $\alpha/\xi$, $1/\beta$ and $\sigma/\xi$ are shown as a function of the number of impurity atoms $N_2$, calculated from Eq.~\eqref{eqn:fj} in (a). The grey dashed line marks where the impurity starts to change the shape of the vortex (see text). Comparisons of the variational wave functions of Eqs.~\eqref{eqn:vpsi1}-\eqref{eqn:vpsi2} with the numerical solutions of the Gross-Pitaevskii equations are shown in (b) and (c) for $N_2=41\text{ and }391$ respectively.}
\end{figure}
\section{\label{sec:nr}Numerical results}
To understand the physical behaviour of the coupled superfluid-impurity system, we perform numerical simulations of the Gross-Pitaevskii model represented by Eqs.~\eqref{eqn:gpe}, subject to the constraint that the phase distribution $\vartheta_1(x,y)$ of the first component satisfies
\begin{equation}\label{eqn:phi1}
\vartheta_1(x,y)=\sum_{j=1}^{N_v}q_j\tan^{-1}\bigg(\frac{y-y_j}{x-x_j}\bigg),
\end{equation}  
which constitutes $N_v$ vortices of charge $q_j$ placed individually at $(x_j,y_j)$. In our work we will take $q_j=\pm 1$. 
Although it is also possible to consider vortices of higher winding number, it is well-known that such configurations are energetically unstable for the Gross-Pitaevskii model represented by Eqs.~\eqref{eqn:gpe}. We work in the healing units, where the healing length is defined as $\xi=\hbar/\sqrt{Mn_{1}g_{\rm 11}}$, the units of energy are given in terms of $[E]=g_{\rm 11}n_1$ and time follows as $[T]=\hbar/g_{\rm 11}n_1$. The wave functions are scaled as $\psi_1\rightarrow\sqrt{n_{1}^{(0)}}\psi_1$ and $\psi_2\rightarrow\psi_2/\xi$. 
\subsection{Single vortices}
Figure \ref{fig:single} presents solutions of the Gross-Pitaevskii model represented by Eqs.~\eqref{eqn:gpe} subject to the phase constraint of Eq.~\eqref{eqn:phi1} for a single vortex with $q_1=+1$ (viz. Eq.~\eqref{eqn:phi1}). Throughout we have taken $N_1=4{\times}10^3$. Panels (a) and (b) present cross-sections of the density of the vortex-carrying superfluid $|\psi_1|^2$ and the impurity $|\psi_2|^2$ as a function of the number of impurity atoms $N_2$ respectively. The effect of changing the number of impurity atoms is quite dramatic -- the shape of the vortex core is observed to change from the characteristic funnel shape presented in panel (i) showing the density and phase for $N_2=1$ to a barrel-shaped hole in the fluid, as presented in panel (ii) for $N_2=391$. Then, individual density cross-sections are presented in panel (c) for the superfluid and impurity components. One can see the qualitative difference between the vortices' core shape (red dash-dotted) for $N_1=4{\times}10^3$ and $N_2=391$ (blue dashed). The corresponding impurity wave functions are plotted in the corresponding colors and styles; while the individual impurity densities are displayed in panels (d-f) for $N_2=1,41,391$. The final panel Fig.~\ref{fig:single} (g) shows a calculation of the width (standard deviation) of the vortex core (black solid) and the impurity wave function (grey dashed).
\subsection{\label{sec:vc}Variational calculation}
To build intuition and explore the physics of the vortex impurity model, we introduce a variational approach to understand the static properties of the mass-imbalanced system. As such we introduce three variational parameters $\alpha$, $\beta$ and $\sigma$ which can be physically interpreted as the vortex length scale, inverse vortex radius and impurity length scale, respectively. The choice of variational ansatz for the superfluid component can be thought of as a hybridization between a logistic and gaussian functions, which as we will demonstrate allows one to smoothly interpolate between a standard superfluid vortex-core profile and the unusual barrel-shaped configuration obtained for large numbers of impurity atoms presented in Fig.~\ref{fig:single}. Then a general purpose variational ansatz (centered at the origin) describing the coupling between the two components $\psi_{1,2}^{\rm var}$ for a vortex of charge $q$ with varying mass imbalance $N_2/N_1$ will be taken as
\begin{widetext}
\begin{subequations}
\begin{align}
\label{eqn:vpsi1}\psi_{1}^{\rm var}(x,y)&=\sqrt{\frac{N_c\beta^2(\beta+1)}{\pi\alpha^2\big[(\beta-1)\ln(\frac{\beta}{\beta+1})+1\big]}}\bigg[\frac{1}{\beta+\exp(-(x^2+y^2)/\alpha^2)}-\frac{1}{\beta+1}\bigg]\exp\bigg[iq\tan^{-1}\bigg(\frac{y}{x}\bigg)\bigg],\\
\label{eqn:vpsi2}\psi_{2}^{\rm var}(x,y)&=\sqrt{\frac{N_2}{\pi\sigma^2}}e^{-(x^2+y^2)/2\sigma^2}.
\end{align}
\end{subequations}
\end{widetext}
The ansatz given by Eqs.~\eqref{eqn:vpsi1} and \eqref{eqn:vpsi2} will be used in what follows to compute the variational energy of the system, as well as to obtain an approximate analytical expression for the angular momentum of the vortex. In order to compare the values of the energy obtained numerically from the Gross-Pitaevskii model represented by Eqs.~\eqref{eqn:gpe} and the renormalized energy of Eqs.~\eqref{eqn:ekin}-\eqref{eqn:vdw2}, we use the superfluid norm $N_c=\int d{\bf r}\big(n_{1}^{(0)} - |\psi_{1}^{\rm var}|^2\big)$ which consequentially allows us to perform a semi-analytic computation of the energy. Proceeding, we can substitute Eqs.~\eqref{eqn:vpsi1}-\eqref{eqn:vpsi2} into Eqs.~\eqref{eqn:ekin}-\eqref{eqn:vdw2} to yield the individual contributions to the total variational energy $E_{\rm v+i}^{\rm var}$, yielding the semi-analytic expressions
\begin{widetext}
\begin{subequations}
\begin{align}
\label{eqn:ven1}E_{\rm1,kin}^{\rm var}&=N_c\frac{\hbar^2}{3M\alpha^2}\frac{\beta+1}{(\beta-1)\ln\big(\frac{\beta}{\beta+1}\big)+1}\bigg\{\ln\bigg(\frac{\beta+1}{\beta}\bigg){-}\frac{\beta}{(\beta+1)^2}{+}3\beta^2q^2\int_{\xi_v/\alpha}^{R/\alpha}\frac{du}{u}\bigg[\frac{1}{\beta+\exp(-u^2)}{-}\frac{1}{\beta+1}\bigg]^2\bigg\},\\
\label{eqn:ven2}E_{\rm 2,kin}^{\rm var}&=N_2\frac{\hbar^2}{2M\sigma^2},\\
\label{eqn:ven3}E_{\rm 11,vdW}^{\rm var}&=N_{c}^{2}\frac{g_{\rm 11}}{12\pi\alpha^2}\frac{6(\beta+1)(\beta-1)^2\ln(\frac{\beta+1}{\beta})+1+3\beta(3-2\beta)}{(\beta+1)\big[(\beta-1)\ln(\frac{\beta}{\beta+1})+1\big]^2},\\
\label{eqn:ven4}E_{\rm 12,vdW}^{\rm var}&=\frac{2g_{\rm 12}N_cN_2}{\pi\alpha^2}\frac{\beta^2(\beta+1)}{(\beta-1)\ln(\frac{\beta}{\beta+1})+1}\int_{0}^{\infty}du u\bigg[\bigg(\frac{1}{\beta+\exp(-\sigma^2u^2/\alpha^2)} - \frac{1}{\beta+1}\bigg)^2-\frac{1}{\beta^2(\beta+1)^2}\bigg]e^{-u^2}.
\end{align}
\end{subequations}
\end{widetext}
To evaluate the various integrals that contribute to the total energy (Eqs.~\eqref{eqn:ekin}-\eqref{eqn:vdw2}), we can exploit the cylindrical symmetry of the problem and evaluate the energy in polar coordinates. The integrals required to obtain these results are listed in appendix \ref{app:ints}. Note that the limits of integration (arising from the vortex phase) in Eq.~\eqref{eqn:ven1} require a little care. The lower limit is the (scaled) length scale of the vortex, the healing length $\xi_v$. This quantity can be approximated from the chemical potential of the homogeneous system, $\mu^{\rm var}_{1}=g_{\rm 11}n_{1}^{(0)}+g_{\rm 12}n_{2}^{(0)}$ where $n_{j}^{(0)}$ is the homogeneous density of each component which can be computed using the variational wave functions Eqs.~\eqref{eqn:vpsi1} and \eqref{eqn:vpsi2}. Then $\xi_v=\hbar/\sqrt{2M\mu_{1}^{\rm var}}$. The upper limit ($\propto R$) of this integral is simply the `radius' of this system, which can be obtained directly for a given simulation using $R=\sqrt{A_{\perp}/\pi}$ where $A_\perp$ is the area of the two-dimensional numerical box. We wish to compare the variational solutions for the energy, Eqs.~\eqref{eqn:ven1}-\eqref{eqn:ven4} to the exact values computed using Eqs.~\eqref{eqn:ekin}-\eqref{eqn:vdw2} and the numerical solutions to the Gross-Pitaevskii model. We use the least squares method to procure the variational parameters $\alpha,\beta$ and $\sigma$ for a given number of atoms $(N_1,N_2)$ and interaction strengths $g_{\rm 11},g_{\rm 12}$. As such we wish to calculate
\begin{figure}[t]
\includegraphics[width=0.95\columnwidth]{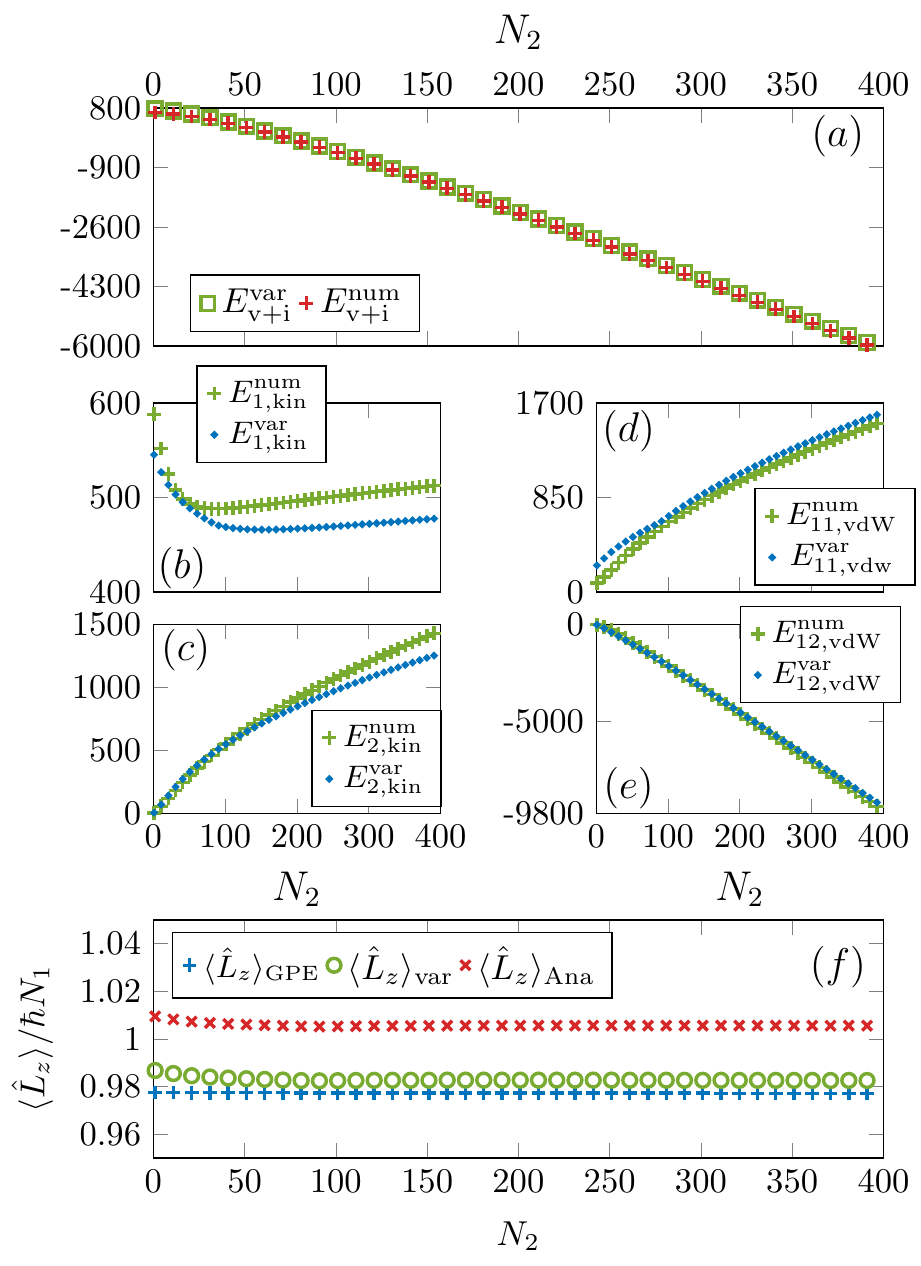}
\caption{\label{fig:varen}(color online) Variational energy and angular momentum. The total variational energy and the total numerical energy are shown as a function of the number of impurity atoms $N_2$ in (a), while the different contributions to the variational energy, Eqs.~\eqref{eqn:ven1}-\eqref{eqn:ven4} are shown in (b-e). The angular momentum $\langle\hat{L}_z\rangle$ is compared in panel (f), computed in three different ways.} 
\end{figure}
\begin{equation}\label{eqn:fj}
\mathcal{F}_j=\underset{\alpha,\beta,\sigma}{\rm min}\sum_{k,l}\norm{|\psi_{j,\alpha,\beta,\sigma}^{\rm var}(x_k,y_l)|^2-|\psi_{j}^{\rm num}(x_k,y_l)|^2},
\end{equation}
where $\psi_{j}^{\rm num}(x,y)$ are the numerical solutions obtained from the Gross-Pitaevskii model represented by Eqs.~(\ref{eqn:gpe}). Equation \eqref{eqn:fj} can then be used to obtain a set of variational parameters for particular atom numbers and scattering lengths. 

Figure \ref{fig:fit} presents the solutions for $\alpha/\xi, 1/\beta$ and $\sigma/\xi$ obtained from Eq.~\eqref{eqn:fj}. The first two of these are the vortex variational parameters, plotted as $\alpha/\xi$ (blue crosses) and $1/\beta$ (green triangles). One can see that the vortex `radius' $1/\beta$ is less for small concentrations of impurity atoms, and increases approximately linearly for larger numbers of impurities. The vortex length scale $\alpha/\xi$ on the other hand increases for small concentrations of impurities, before leveling off at larger impurity numbers. There appears to be some correlation between the vortex length scale $\alpha/\xi$ and radius $1/\beta$, which seems to occur for $N_2\simeq100$, i.e. at the point where the length scale $\alpha/\xi$ levels off and the radius $1/\beta$ starts to grow (grey dashed line). A possible explanation for this is that at this concentration of impurity atoms the shape of the vortex starts to significantly change, developing the flat-bottomed profile (Fig.~\ref{fig:single}(c)), leading in turn to competition between the two terms appearing in the first fraction in square brackets in Eq.~\eqref{eqn:vpsi1}. The impurity length scale $\sigma/\xi$ is also shown (open red circles). An accompanying fit is displayed (black dashed) to $\sigma_{\rm fit}/\xi\propto\sqrt{N_2}$, which can be inferred from Eq.~\eqref{eqn:ven2}, showing good agreement.

The expressions for the variational energy, Eqs.~\eqref{eqn:ven1}-\eqref{eqn:ven4} can be compared to the exact values found from the numerical solutions of the Gross-Pitaevskii model represented by Eqs.~\eqref{eqn:gpe} and the renormalized energy of Eqs.~\eqref{eqn:ekin}-\eqref{eqn:vdw2}. This is presented in figure \ref{fig:varen}. Panel (a) shows a comparison of the total variational energy (green squares) and the total energy calculated from the exact numerical solutions (red crosses) for $\psi_{\rm 1,2}$ as a function of the number of impurity atoms, showing good agreement. Then, panels (b-e) show the individual contributions to the total energy, while the two kinetic terms for the vortex and impurity, $E_{\rm j,kin}$ are shown in (b) and (c), respectively, and the two interaction energies $E_{\rm 11,vdW}$ and $E_{\rm 12,vdW}$ are shown in (d) and (e), respectively. In general there seems to be good agreement between the two energies amongst the various contributions to the total energy, with the exception of the kinetic energy contribution, panel (b). This small difference likely originates from approximating the length scale $\xi_v$ of the vortex using the chemical potential of the homogeneous system.

The variational approach can also be used to calculate an expression for the angular momentum of the superfluid system. The expectation value of the $z$-projection of the angular momentum is defined as $\langle\hat{L}_z\rangle=\int d{\bf r}\ \psi_{\rm var}^{*}\hat{L}_z\psi_{\rm var}$. Using the representation $\psi_{\rm var}=\sqrt{n_{\rm var}}\exp(i\vartheta_{\rm var})$ to separate the density $n_{\rm var}$ and phase $\vartheta_{\rm var}$ variables, the angular momentum can then be computed asymptotically using Eq.~\eqref{eqn:vpsi1} and the assumption $R/\alpha\gtrsim1$ as
\begin{equation}
\label{eqn:ama}\frac{\langle\hat{L}_z\rangle_{\rm var}}{\hbar}=N_c\frac{R^2/\alpha^2-(\beta+1)+(\beta^2-1)\ln(\frac{\beta+1}{\beta})}{(\beta+1)\big[(\beta-1)\ln(\frac{\beta}{\beta+1})+1\big]}.
\end{equation}  
The angular momentum of $\psi_1$ is calculated and compared in panel (f) of Fig.~\ref{fig:varen} with three different approaches. The first two data sets show $\langle\hat{L}_z\rangle$ computed using the Gross-Piteavskii solutions (blue plusses) and the variational angular momentum $\langle\hat{L}_z\rangle_{\rm var}$, Eq.~\eqref{eqn:ama} (green circles). We can see that all three approaches are almost independent of $N_2$, as we would expect, and are only a few percent away from the theoretical value $\langle\hat{L}_z\rangle_{\rm Ana}=\hbar N_1$ (red crosses).
\begin{figure}[t]
\includegraphics[width=0.9\columnwidth]{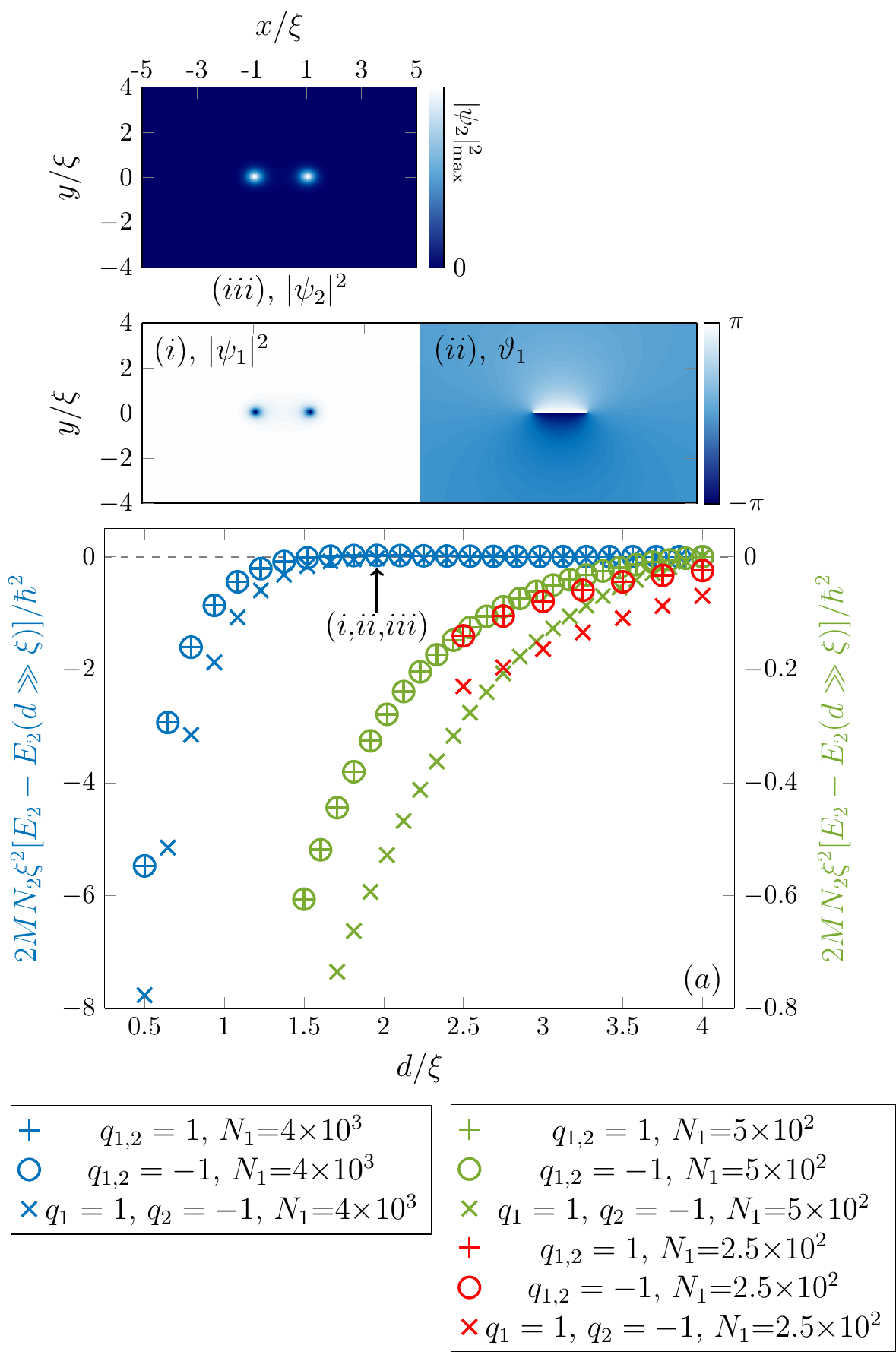}
\caption{\label{fig:vp}(color online) Vortex pair energies. Panel (a) shows the spectral flow (eigenvalues) $E_2-E_2(d\gg\xi)$ of the impurity for different combinations of vortex charges (Eq.~\eqref{eqn:vpp}) and mass imbalances for a single impurity $N_2=1$. Examples of the stationary states are shown in panels (i)-(iii) corresponding to a vortex dipole with $N_1=4{\times}10^{3}$ (see black arrow, (a)). Note that the left-axis data in (a) correspond to the left legend, while the right-axis data correspond to the right legend.}
\end{figure}
\subsection{Vortex pairs}      
As well as individual vortices, it is also useful to consider pairs of vortices; either with the same sign $(q_{\rm 1,2}=\pm1)$ or with differing signs $(q_{\rm 1}=1,q_{\rm 2}=-1)$ (see Eq.~\eqref{eqn:phi1}). Then the resulting phase distribution for the vortex pair is
\begin{equation}\label{eqn:vpp}
\vartheta_{\rm 1}(x,y)=q_{\rm 1}\tan^{-1}\bigg[\frac{y}{x{+}d/2}\bigg]+q_{\rm 2}\tan^{-1}\bigg[\frac{y}{x{-}d/2}\bigg],
\end{equation}
where the vortices are separated by a distance $x=d$. By changing the separation $d$ and mass imbalance ratio $N_2/N_1$, we can explore the effect that impurity atoms have. Our analysis is based on a static vortex configuration, hence it is instructive to consider when such an analysis is valid in the dynamical regime. A pair of like or opposite-sign vortices will co-rotate or propagate linearly respectively. The angular frequency or velocity at which they do this depends in both cases on their relative separation. For two vortices placed in close proximity ($d\sim\xi$), their motion will be faster relative to a larger separation ($d\gg\xi$). In this work it is this second regime that we will consider -- hence additional contributions to the impurities kinetic energy from the dynamics will be mitigated in this limit, and our analysis based on Eq.~\eqref{eqn:vpp} is appropriate. 
\begin{figure*}[t]
\includegraphics[scale=1]{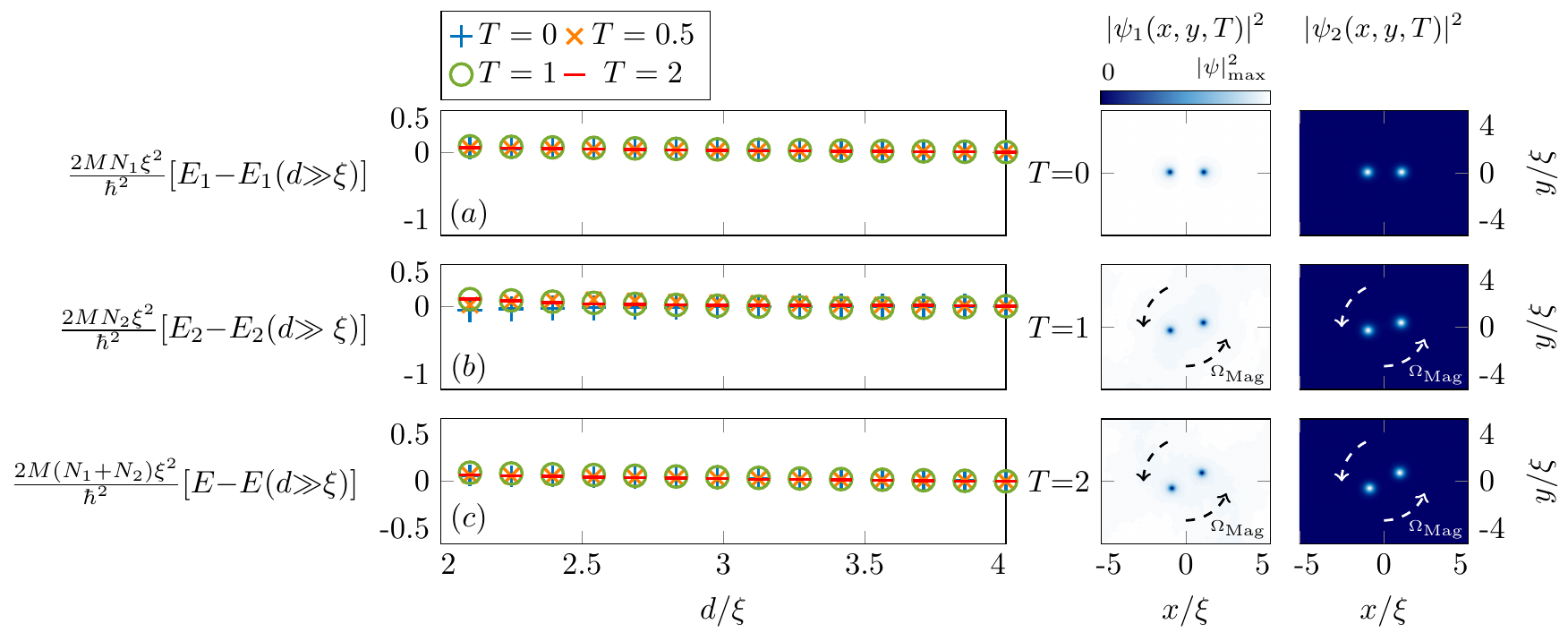}
\caption{\label{fig:dyn}(color online) Vortex-impurity dynamics. Panels (a-c) show the energies $E_1$, $E_2$ and the total energy $E$ as a function of displacement $d/\xi$ for $T=0,0.5,1,2$. The six accompanying density plots $|\psi(x,y,T)|^2$ show examples of dynamics for the case of co-rotation, $q_1=q_2=+1$ with $d/\xi=2$.}
\end{figure*}

Figure \ref{fig:vp} shows the computed impurity energies $E_2=\int d{\bf r}\psi_{2}^{*}[-(\hbar^2/2M)\nabla^2+g_{\rm 12}|\psi_1|^2]\psi_{2}/N_2$ for different vortex charges and mass imbalances as a function of vortex separation. Note that the asymptotic energy $E_2(d\gg\xi)$ (i.e. the energy of two well separated vortices) has been subtracted for a given atom number $N_1$ to aid comparison. At large mass imbalances $N_2/N_1=1/4{\times}10^{3}$ there is a small splitting between the like-sign and vortex dipole energies (blue data, (a)) while for larger imbalances the splitting is more pronounced (e.g. $N_2/N_1=1/500$ (green data, (a)) and $N_2/N_1=1/250$ (red data)). We attribute this difference to the short-range nature of the vortices densities, which for a vortex dipole manifest as a depression when the vortices approach, resulting in a lower impurity energy than for a pair of like-sign vortices. Then, the three panels $(i)$-$(iii)$ show example stationary states obtained from the solution of the Gross-Pitaevskii model represented by Eqs.~\eqref{eqn:gpe}. Here $(i)$ and $(ii)$ show the density $|\psi_1|^2$ and phase $\vartheta_1$ of the first component,  respectively; while $(iii)$ shows the delocalized impurity density $|\psi_2|^2$. These calculations reveal a simple relationship between the mass imbalance $N_2/N_1$ and the splitting of the impurities energy, with smaller mass imbalances leading to larger differences in the impurities energy at short separations. Since we know that a vortex pairs energy is logarithmic in their separation $d$ such that $E_{\rm 12}\propto q_1q_2\ln(R/d)$, larger vortices will comparatively feel an increased repulsion or attraction at the same distance $d$ than those with smaller core sizes, consistent with the known behaviour of vortices. Figure \ref{fig:vp} demonstrates that the presence of impurities will contribute an \textit{attractive} interaction between vortices.

To investigate the dynamical behaviour of the energy $E_2$ for $T>0$, we can evolve the Gross-Pitaevskii model represented by Eqs.~\eqref{eqn:gpe} in time to some later time $T$, and re-calculate the energies using the new $\psi_{\rm 1,2}(x,y,T)$ of the superfluid and impurity components. In Fig.~\ref{fig:dyn} the energies $E_1(T)$, $E_2(T)$ and $E(T)$ of vortex pairs are presented for $N_1=4{\times}10^{3}$, $N_2=1$ with vortex charges $q_1=q_2=+1$ for $T=0,0.5,1,2$ in panels (a-c) respectively. Accompanying this are six examples of the vortices and impurities densities at times $T=0,1,2$. The vortices displacement is $d=2\xi$. Then for $T>0$ the anti-clockwise rotation direction is indicated by $\Omega_{\rm Mag}$. Panel (b) shows the important time-dependent impurity energies, $2MN_2\xi[E_2-E_2(d{\gg}\xi)]/\hbar^2$ for separations in the range $d/\xi$=$2:4$. Already even modest separations show that this quantity does not change appreciably during dynamics, while at larger separations there is almost no difference to the static case. Complementary to this, since we consider the limit $N_2/N_1\ll1$, the effect of excitations such as sound waves from the impurity on the superfluid component will be minimal, since fluctuations will scale as $\sqrt{N_1N_2}$, which would be sub-leading compared to those originating from interactions between superfluid atoms in the first component. The  calculations presented here support the use of static vortex configurations to investigate the properties of vortex-trapped impurities in this manner.
     
\section{\label{sec:cb}Covalent bonds between well-separated vortices}

In this section we consider the possibility of bonding between a pair of like-sign vortices. Motivated by our findings in the previous section where the interaction between two well-separated vortices with both like and different signs were considered, we consider how the excited states of such a system can contribute an additional attractive force. Note that the covalent force we will study in this section is the next-to-leading order interaction (the leading interaction is ${\sim}\ln(d/\xi)$). Also, the well separated vortices co-rotate or co-move with their separation kept constant, so that the inver-vortex forces do not change appreciably in time. Hence, we will deal with only static vortices in this section.

\subsection{Pad\'e approximation for a single vortex}

\begin{figure}[t]
\includegraphics[scale=0.8]{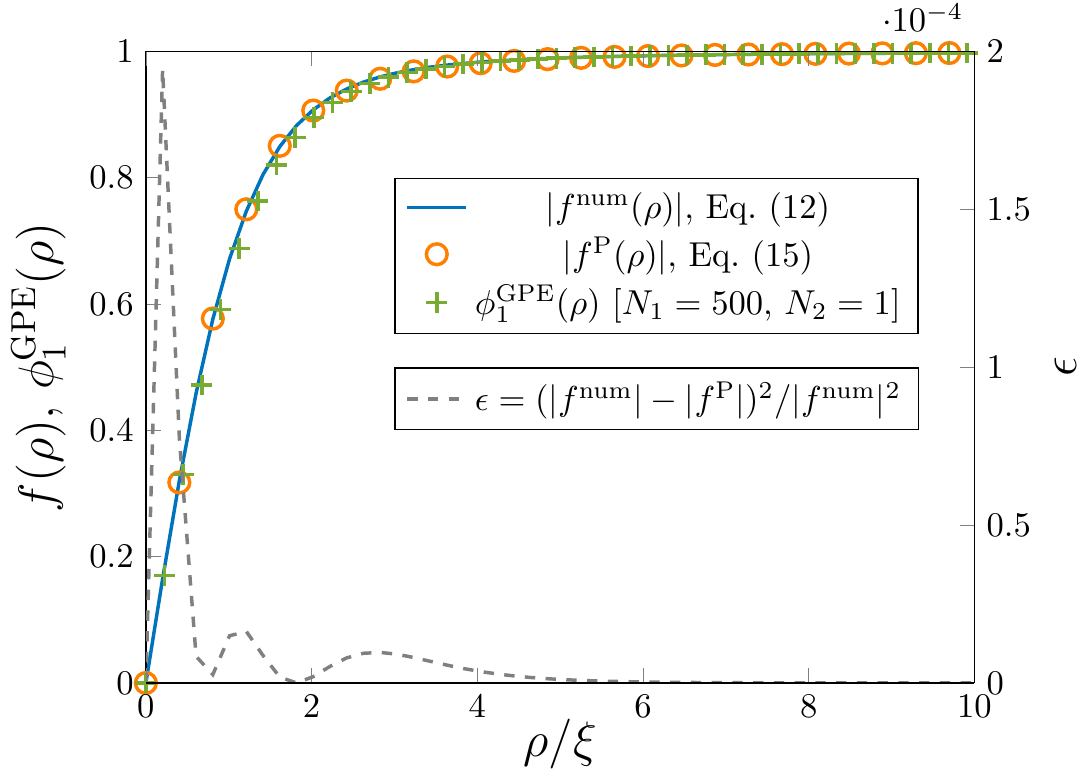}
\caption{\label{fig:pade}(color online) Pad\'e approximation. The vortex solution $|f^{\rm num}(\rho)|$ to Eq.\eqref{eqn:f} (blue solid line) and the accompanying Pad\'e approximation, $|f^{\rm P}(\rho)|$ of Eq.~\eqref{eqn:pade} (open orange circles) are shown (left vertical scale). The scaled solution $\phi^{\rm GPE}_{1}(\rho)$ to Eqs.~\eqref{eq:GP1}-\eqref{eq:GP2} with $N_1=500, N_2=1$ is shown as green crosses. The scaled difference $\epsilon=(|f^{\rm num}|-|f^{\rm P}|)^2/|f^{\rm num}|^2$ (grey dashed line) is shown on the right vertical scale.}
\end{figure}

Let us begin by writing the time-independent form of the Gross-Pitaevskii model represented by Eqs.~\eqref{eqn:gpe} using the transformation $\psi_j({\bf r},t)=\exp(-iN_jE_j t/\hbar)\phi_j({\bf r})$, giving 
\begin{align}
\label{eq:GP1}\bigg[-\frac{\hbar^2}{2M}\nabla^2+g_{\rm 11}|\phi_1|^2+g_{\rm 12}|\phi_2|^2\bigg]\phi_1=N_1E_1\phi_1, \\
\bigg[-\frac{\hbar^2}{2M}\nabla^2+g_{\rm 12}|\phi_1|^2\bigg]\phi_2=N_2E_2\phi_2.
\label{eq:GP2}
\end{align}
A dimensionless set of units can be adopted by scaling the units of length by the healing length $\xi$, $\phi_1\rightarrow\sqrt{n_1}\phi_1$ and $\phi_2\rightarrow\phi_2/\xi$. We then perform one additional scaling, $\phi_j\rightarrow\sqrt{N_j}\phi_j$ which allows us to understand the parameter regimes in terms of the coupling between the superfluid and the impurity. This yields the coupling parameters (corresponding to the dimensionless coefficients of $|\phi_1|^2$ and $|\phi_2|^2$ in Eq.~\eqref{eq:GP2} and \eqref{eq:GP1}, respectively) as $C_1=2Mg_{\rm 12}\xi^2n_1N_j/\hbar^2$ and $C_2=2Mg_{\rm 12}N_2/\hbar^2$, while the units of energy are scaled as $2MN_j\xi^2E_j/\hbar^2$ for the energy of component $j$. The ratio of coupling constants yields $C_2/C_1=(N_2/N_1)(1/n_1\xi^2)$, which in the limit $N_1\gg N_2$ allows us to decouple Eqs.~\eqref{eq:GP1}-\eqref{eq:GP2}. One can also consider the case of unequal masses $M_1{\neq}M_2$, in which case this condition becomes instead $(M_1N_2/M_2N_1)(1/n_1\xi^2)\ll 1$, allowing additional parameter flexibility in studying the impurity physics of the model of Eq.~\eqref{eq:GP1}-\eqref{eq:GP2}. Then, by separating the phase and amplitude of the vortex as $\phi_1=\sqrt{\frac{N_1E_1}{g_{11}}} e^{iq\theta}f(\rho)$ we obtain a simplified form of Eq.~\eqref{eq:GP1}
\begin{equation}\label{eqn:f}
-\xi^{2}\bigg(\frac{d^2f}{d\rho^2}+\frac{1}{\rho}\frac{df}{d\rho}-\frac{f}{\rho^2}\bigg)+2(f^2-1)f=0,
\end{equation}
To build the Pad\'e approximation, we can solve Eq.~\eqref{eqn:f} in two limits subject to the boundary conditions $f(\rho\rightarrow0)=0$ and $df(\rho)/d\rho|_{\rho\rightarrow\infty}=0$. This gives the pair of asymptotic solutions 
\begin{align}
\label{eqn:phiz}f(\rho\rightarrow0)&=a \rho/\xi+\mathcal{O}\big(\rho^{3}\big),\\
\label{eqn:phii}f(\rho\rightarrow\infty)&=1-\frac{\xi^2}{4\rho^2}+\mathcal{O}\big(\rho^{-4}\big)
\end{align}
here the constant $a$ appearing in Eq.~\eqref{eqn:phiz} is the first fitting constant used to build the Pad\'e approximation \cite{rorai_2013}. We note that the authors of \cite{braz_2020} used a piece-wise function to perform a variational analysis of the vortex's allowed bound states. Such an approach is useful for analytical treatments, however since we wish to obtain the excited states numerically, the Pad\'e approximation provides an accurate approximation to the numerical solution of the Gross-Pitaevskii model, and hence a closer agreement to the true numerical spectrum of the impurity. Then, an appropriate approximation for a single vortex is written as
\begin{eqnarray}\label{eqn:pade}
f^{\rm P}(\rho)=\sqrt{\frac{a^2(\rho/\xi)^2 + 2(b^2-a^2)(\rho/\xi)^4}{1+b^2(\rho/\xi)^2 +2(b^2-a^2)(\rho/\xi)^4}},
\end{eqnarray}
which satisfies both asymptotic boundary conditions given by Eqs.~\eqref{eqn:phiz} and \eqref{eqn:phii}. Figure~\ref{fig:pade} shows a comparison between the numerical solutions to Eq.~\eqref{eqn:f} (blue solid line) and the Pad\'e approximation, Eq.~\eqref{eqn:pade} (open orange circles). The two fitting parameters are found to be $a\sim0.81$ and $b\sim0.88$. Accompanying this, the (scaled) solution $\phi^{\rm GPE}_{1}$ obtained from Eqs.~\eqref{eq:GP1}-\eqref{eq:GP2} with $N_1=500$, $N_2=1$ and $C_2=2Mg_{\rm 12}N_2/\hbar^2=1$ is shown for comparison (left vertical axis). The dashed grey data shows the difference $\epsilon=(|f^{\rm num}|-|f^{\rm P}|)^2/|f^{\rm num}|^2$ (right vertical axis) between the numerical and Pad\'e approximations, which being small, supports this choice.
\subsection{\label{sec:ipm}Impurity potential model}
\begin{figure*}[t]
\includegraphics[width=0.9\textwidth]{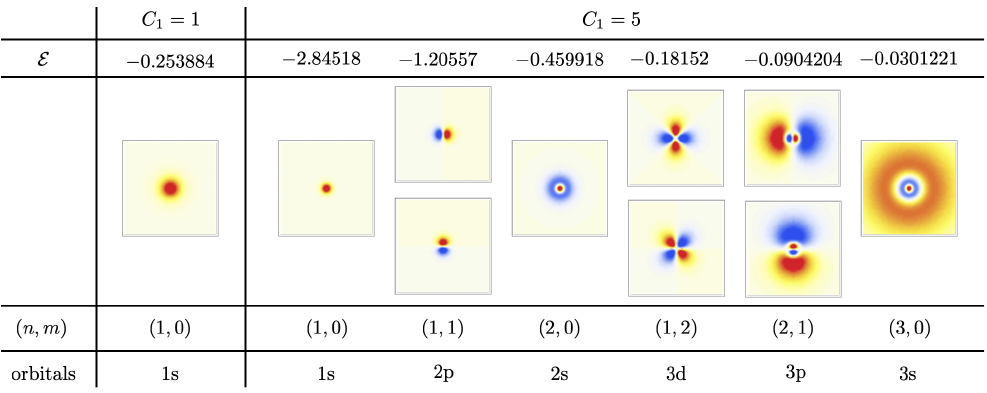}
\caption{(color online) The impurity wave functions $\phi^{\text{1s}}_2$ for $C_1=2Mg_{\rm 12}\xi^2n_1N_1/\hbar^2 = 1$, and $\phi^{\text{1s}}_2$, $\phi^{\text{2p}}_2$, $\phi^{\text{2s}}_2$, $\phi^{\text{3d}}_2$, $\phi^{\text{3p}}_2$, and $\phi^{\text{3s}}_2$ for $C_1=5$ are shown. The red color means $\phi^{m,n}_2 > 0$, while the blue corresponds to $\phi^{m,n}_2<0$, and the white $\phi^{n,m}_2 = 0$. The plot region is $x/\xi \in [-10,10]$ and $y/\xi \in [-10,10]$.}
\label{fig:bs_single}
\end{figure*}
Having the analytic Pad\'e approximation of Eq.~\eqref{eqn:pade} for $f^{P}(\rho)$, we can proceed to solve Eq.~(\ref{eq:GP2}). Since we assume $(N_2/N_1)(1/\xi^2n_1)$ is small, we can treat $\phi_1$ in Eq.~\eqref{eq:GP2} as an external potential by omitting any back reaction between the vortex and the impurity. We are left with the two-dimensional quantum mechanical problem of calculating the bound states of $\phi_2$ of an impurity by the axially symmetric Schr\"odinger potential $g_{\rm 12}f^{P}(\rho)^2$. Then we have
\begin{eqnarray}
\bigg[-\frac{\hbar^2}{2M}\nabla^2+\mathcal{U}(\rho)\bigg]\phi_2=\mathcal{E}\phi_2,
\label{eq:schrodinger}
\end{eqnarray}
here the potential $\mathcal{U}(\rho)=g_{\rm 12}\big(|\phi_{1}^{\rm P}|^2-N_1E_1/g_{\rm 11}\big)$ satisfies $\mathcal{U}(\rho\rightarrow\infty)=0$ while the energy is $\mathcal{E}=N_2E_2-N_1E_1g_{\rm 12}/g_{\rm 11}$. Then, the Schr\"odinger potential $\mathcal{U}(\rho)$ obeys $\mathcal{U}(0) = -N_2E_2g_{\rm 12}/g_{\rm 11}$, $\mathcal{U}(\rho\rightarrow\infty) = 0$, hence the potential exists on the interval $-N_2E_2g_{\rm 12}/g_{\rm 11} \le \mathcal{U}(\rho) < 0$. Since the potential $\mathcal{U}(\rho)$ is axisymmetric, we can separate $\rho$ and $\theta$ as
\begin{equation}
\phi_2(\rho,\theta) = \sqrt{\frac{N_2E_2}{g_{\rm 11}}}\frac{\varphi(\rho)}{\sqrt{\rho}}\exp(im\theta),
\label{eq:varphi}
\end{equation}
here $m \in \mathbb{Z}$ defines the quantum number of angular momentum for the impurity, with the associated operator $\hat{L}_z=-i\partial_\theta$. Normalization for $\varphi(\rho)$ is then imposed through $2\pi\int^{\infty}_{0} d\rho\ \varphi(\rho)^2=N_2$. Inserting Eq.~(\ref{eq:varphi}) into Eq.~(\ref{eq:schrodinger}), we are left with the following one-dimensional Schr\"odinger problem
\begin{eqnarray}
\left[\hat{Q}^\dagger \hat{Q} + m^2\frac{\hbar^2}{2M\rho^2} + \mathcal{U}(\rho)\right] \varphi_{n,m} = {\cal E}_{n,m} \varphi_{n,m},
\label{eq:schrodinger_2}
\end{eqnarray}
where we have introduced the operator
\begin{eqnarray}
\hat{Q} = \frac{\hbar}{\sqrt{2M}}\left(-\frac{d}{d\rho} + \frac{1}{2\rho}\right),
\end{eqnarray}
then we have $\hat{Q}^\dagger\hat{Q}=-(\hbar^2/2M)(d^2/d\rho^2+1/4\rho^2)$, while the term $m^2\hbar^2/2M\rho^2$ in Eq.~(\ref{eq:schrodinger_2}) defines the centrifugal potential felt by the impurity. To understand the basic properties of the impurity, let us first consider $m=0$. Note that the operator $\hat{Q}^\dagger\hat{Q}$ is positive semi-definite. Therefore, when $\mathcal{U}(\rho)$ vanishes ($g_{\rm 12}\rightarrow0$), the eigenvalue $\mathcal{E}$ satisfies ${\cal E}\big|_{g_{\rm 12} \rightarrow 0} \ge 0$. The lowest eigenstate $(n=1)$ has ${\cal E} = 0$ and should satisfy $\hat{Q}\varphi = 0$. This can be easily solved by $\varphi(\rho) \propto \sqrt \rho$. This in turn corresponds to constant $\phi_2$, which is clearly non-normalizable. Now we consider small but finite $g_{\rm 12}$ by slightly increasing $g_{\rm 12}$ from zero. This has the effect of lowering the Schr\"odinger potential $\mathcal{U}(\rho)$ everywhere since $\mathcal{U}(\rho)<0$. Then, because we can continuously change $g_{\rm 12}$ from $0$ to $g_{\rm 12}>0$, there is a corresponding continuous shift of the eigenvalues. Namely, the ground state energy should satisfy ${\cal E}_{1,0} < 0$ for $g_{\rm 12} \neq 0$. Therefore, the Schr\"odinger equation at $\rho \to \infty$ reads $d^2\varphi_{1,0}/d\rho^2= 2M|{\cal E}_{1,0}|\varphi_{1,0}/\hbar^2$, the solution for which is a bound state whose asymptotic wave function is 
\begin{equation}
\varphi_{1,0}(\rho)\propto\exp\bigg(-\sqrt{2M|{\cal E}_{1,0}|}\,\rho/\hbar\bigg).
\end{equation}

To summarize, we have shown that there exists at least one bound state regardless of the form of the potential $\mathcal{U(\rho)}$. On the other hand, the existence of (bound) excited states $\varphi_{n,0}(\rho)$ for $n>0$ sensitively depends on $g_{\rm 12}$. In general the greater $g_{\rm 12}$ is, the more excited states exist and hence there are more possibilities to form bound states with a particular potential. On the other hand, for $m\neq0$, the centrifugal potential always lifts the Schr\"odinger potential since it contributes a positive energy to the total potential felt by the impurity. Therefore, a sufficiently large $g_{\rm 12}$ is required to accommodate bound states for $\varphi_{n,0}$ ($n>1$).

\subsection{\label{sec:sbs}Single vortex bound states}

Let us next numerically obtain the eigenvalues. Having the analytic approximation $f^{\rm P}(\rho)$ in hand is an advantage for this purpose. 
We choose two examples $C_1=2Mg_{\rm 12}\xi^2n_1N_1/\hbar^2 = 1$ and $C_1=5$ as demonstrations. For the small $C_1=1$, there exists only one ground state $(n,m) = (1,0)$ with the energy $2MN_2\xi^2{\cal E}_{1,0}/\hbar^2 \simeq -0.25$. We show a two-dimensional plot of $\phi^{1,0}_2$ in Fig.~\ref{fig:bs_single}. For the large $C_1=5$ the potential is deeper, so that we have six bound states with $(n,m) = (1,0), (1,1), (2,0), (1,2), (2,1), (3,0)$. The corresponding energy eigenvalues $2MN_2\xi^2{\cal E}_{n,m}/\hbar^2$ are $-2.85$, $-1.21$, $-0.46$, $-0.18$, $-0.09$, and $-0.03$, respectively. The corresponding number of circular nodes is given by $n-1$ and that of linear nodes is $|m|$. Then, the total number of nodes is given by $N = n + |m| -1$. Therefore, $(1,1)$ and $(2,0)$ have a single node, and we observe ${\cal E}_{1,1} < {\cal E}_{2,0}$. Similarly, $(1,2)$, $(2,1)$ and $(3,0)$ have double nodes, and the corresponding energies obey ${\cal E}_{1,2} < {\cal E}_{2,1} < {\cal E}_{3,0}$. From these observations, we see that creating a circular node requires more energy than a linear one. Fig.~\ref{fig:bs_single} shows $\phi_{n,m}$ for the six bound states. We adopt the conventional names for atomic orbitals. For example, the bound state $(n,m)= (1,0)$ is called the 1s orbital because the first number 1 corresponds to $N+1=1$ while s is for $|m|=0$.  
Then we can adopt the conventional rules, (1,1)=2p, (2,0)=2s, (1,2)=3d, (2,1)=3p, (3,0)=3s, etc. Note that, by increasing $2Mg_{\rm 12}\xi^2n_1N_1/\hbar^2$ from $1\to5$, the potential depth becomes deeper while its width is unchanged. Accordingly, the wave functions for $C_1=5$ are more compact than those of $C_1=1$. We note that the authors of \cite{braz_2020} also studied the excited states of a single impurity confined by a vortex; however our approach differs somewhat since we will in the remaining subsections demonstrate how these states can be used to build an analogy with chemical bonds.
\begin{figure}[t]
\begin{center}
\includegraphics[width=0.9\columnwidth]{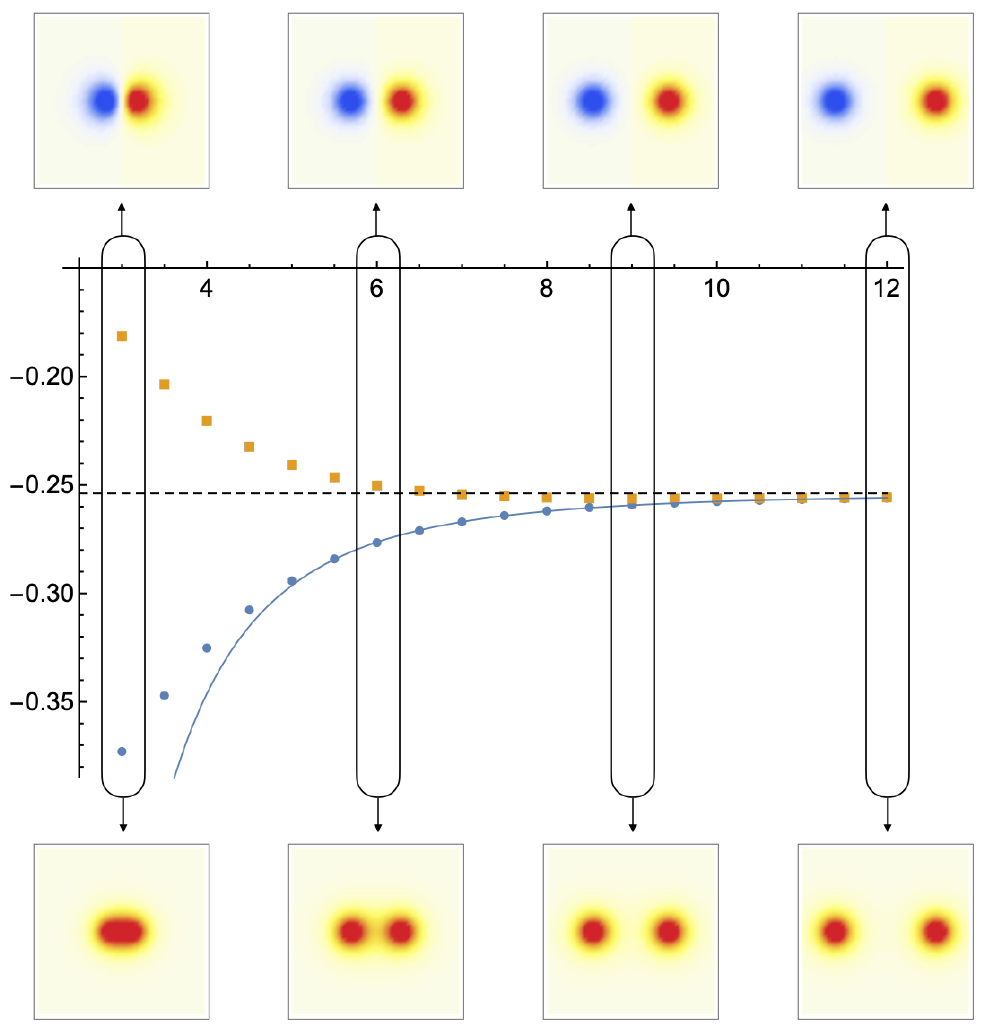}
\caption{(color online) The spectral flow of 1s-1s orbitals for $C_1=2Mg_{\rm 12}\xi^2n_1N_1/\hbar^2=1$. The horizontal axis corresponds to the dimensionless distance $d/\xi$ while the vertical axis is the corresponding dimensionless energy eigenvalue $2MN_2\xi^2{\cal E}/\hbar^2$. The dotted horizontal line shows ${\cal E}_{\text{1s}}^{(1)}$, and the blue solid curve is a numerical fit shown in Eq.~(\ref{eq:numerical_fit}). The blue disks are for the states which are even (no nodes) for $x/\xi \to -x/\xi$, while the orange squares are for those which have odd parity. We show the wave functions of the ground states for $d/\xi=3,6,9,12$ on the bottom line, and those of the first excited state on the top line. The spatial region of the density plots is $x/\xi \in [-10,10]$ and $y/\xi \in [-10,10]$.}
\label{fig:covalent_1}
\end{center}
\end{figure}
\subsection{Covalent bonds between two vortices}

In this subsection we compute the excited states of two neighboring vortices. We consider two vortices placed at $(x,y) = (\pm d/2,0)$. For the background configuration of $\phi_1$, we take a product Ansatz
\begin{eqnarray}\label{eqn:vprod}
\phi_1^{\rm(2)}(x,y,d) {=} \sqrt{\frac{g_{\rm 11}}{N_1E_1}}f^{\rm P}\left(x{-}\frac{d}{2},y\right)f^{\rm P}\left(x{+}\frac{d}{2},y\right),
\end{eqnarray}
and we solve Eq.~\eqref{eq:schrodinger} with the replacement $\mathcal{U}(\rho)\rightarrow g_{\rm 12}(|\phi_{1}^{(2)}|^2-N_1E_1/g_{\rm 11})$. The presence of one vortex affects the other vortex, and the distortion cannot be ignored when the separation $d$ is small. Therefore, in what follows we will take $d\gg\xi$ for the above product Ansatz to be a valid approximation. As explained in Sec.~\ref{sec:ipm}, we omit any back reaction between the field $\phi_1$ and $\phi_2$ by keeping $(N_2/N_1)(1/\xi^2n_1)\ll1$.
\begin{figure}[b]
\begin{center}
\includegraphics[width=\columnwidth]{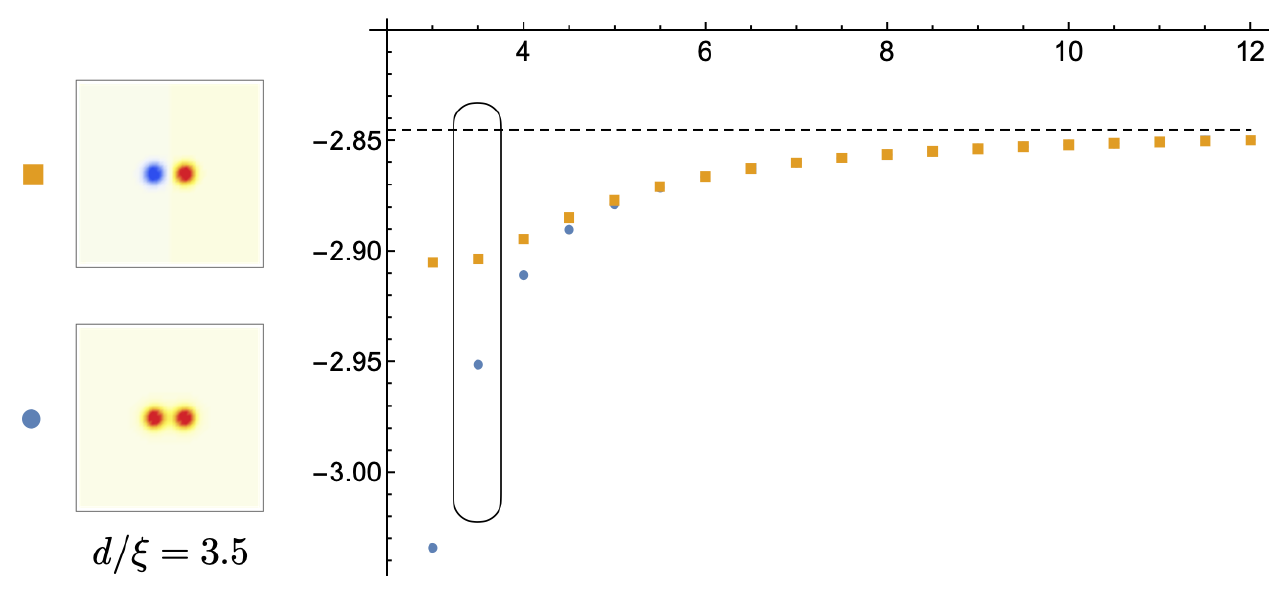}
\caption{(color online) The spectral flow of 1s-1s orbitals for $2Mg_{\rm 12}\xi^2n_1N_1/\hbar^2=5$. The horizontal axis corresponds to the dimensionless distance $d/\xi$ while the vertical axis is the corresponding dimensionless energy eigenvalue $2MN_2\xi^2{\cal E}/\hbar^2$. The dotted horizontal line shows the energy of the state ${\cal E}_{\text{1s}}^{(1)}$. The blue disks correspond to states which are even (no nodes) for $x/\xi \to -x/\xi$, while the orange square are for those which have odd parity. We show the two wave functions for $d/\xi=3.5$. The spatial region of the density plots is $x/\xi \in [-10,10]$ and $y/\xi \in [-10,10]$.}
\label{fig:bs_double_10}
\end{center}
\end{figure}
In the previous subsection \ref{sec:sbs}, it was explained that each background $\phi_1$ vortex represents an attractive potential for the impurity atoms. Therefore, when we place two such vortices on the plane, the impurities are attracted by such a pair of vortices. However, when the displacement $d\gg\xi$, the individual vortex wave functions no longer overlap, so the energy spectra are the same as those of a single vortex, $2MN_2\xi^2{\cal E}_{\text{1s}}^{(1)}/\hbar^2 \simeq -0.25$. On the other hand, when the vortex separation $d$ is large but finite, the degeneracy is lifted and instead the spectra are slightly split. Since the problem is no longer axially symmetric, we consider instead Eq.~(\ref{eq:GP2}) with $\phi_1$ replaced by $\phi_1^{(2)}$, from which we obtain the numerical solutions. 
\begin{figure*}[t]
\begin{center}
\includegraphics[width=0.7\textwidth]{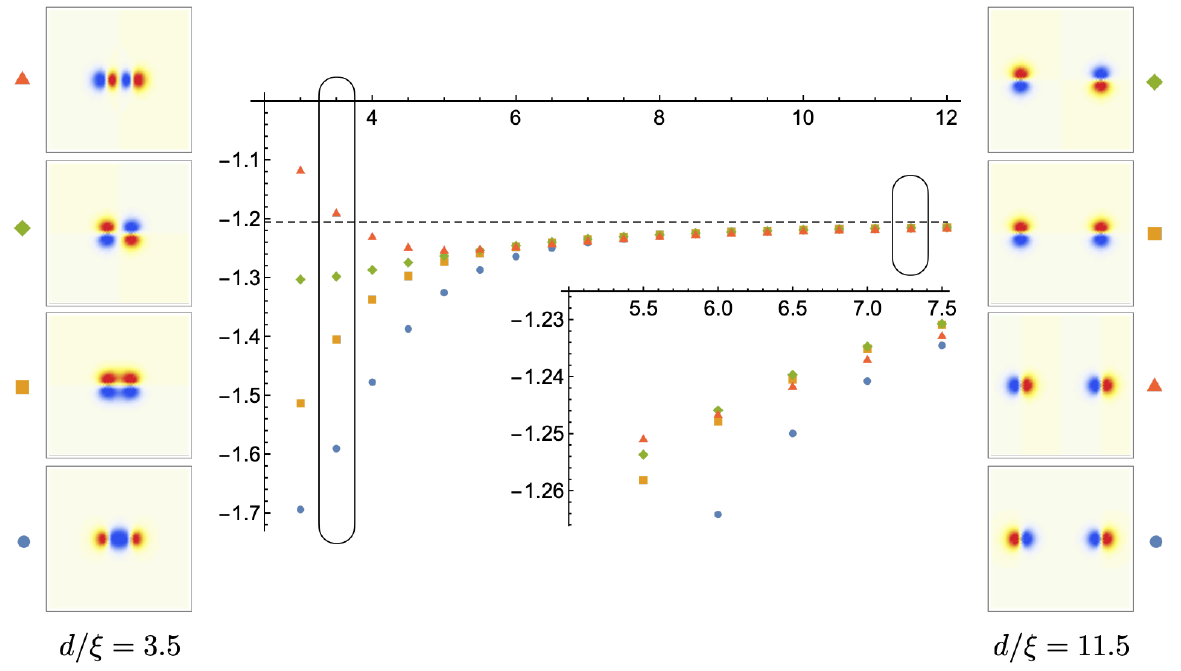}
\caption{(color online) The spectral flow of 2p-2p orbitals for $2Mg_{\rm 12}\xi^2n_1N_1/\hbar^2=5$. The horizontal axis corresponds to the dimensionless distance $d/\xi$ while the vertical axis corresponds to the dimensionless energy eigenvalue $2MN_2\xi^2{\cal E}/\hbar^2$. The dotted horizontal line shows the energy for the state ${\cal E}_{\text{2p}}^{(1)}$. The blue disks correspond to the $|\mathlarger{\mathlarger{\circ}}\rangle$ states (Eq.~\eqref{eqn:circ}), $|\triangle\rangle$ states (orange triangles, Eq.~\eqref{eqn:tri}), $|\square\rangle$ states (yellow squares, Eq.~\eqref{eqn:sq}), and the $|\diamond\rangle$ states are represented as green diamonds (Eq.~\eqref{eqn:dia}). We show the wave functions for $d=11.5\xi$ in the right-most column, and those of $d=3.5\xi$ in the left-most column. The spatial region of the density plots is $x/\xi \in [-10,10]$ and $y/\xi \in [-10,10]$.}
\label{fig:bs_double_11}
\end{center}
\end{figure*}
Let us start with the simple case of $C_1=1$ for which each vortex can have only a single 1s orbital. We vary $d/\xi$ from $d=3\xi$ to $d=12\xi$ with step size $\delta d = \xi/2$. For each given $d$, we numerically solve Eq.~(\ref{eq:GP2}) and determine the energy eigenvalues. The results are summarized in Fig.~\ref{fig:covalent_1}. As expected we have two states, a state with even symmetry, while the other is an odd function for the parity $x/\xi \to -x/\xi$. The former is the ground state whereas the latter is the excited state, since it has one linear node on the $y$ axis. The two states are almost degenerate when $d$ is sufficiently large (see $d \gtrsim 10\xi$ of Fig.~\ref{fig:covalent_1}). These states lie slightly below the ground state energy of the single vortex, ${\cal E}_{\text{1s}}^{(1)}$, and for large but finite $d$, they asymptotically approach ${\cal E}_{\text{1s}}^{(1)}$ from below. Therefore, at asymptotically large $d$, having impurities for both the ground state and the excited state results in an attractive force between the two background vortices of $\phi_1$. By decreasing $d\,(\lesssim 9\xi)$, however, the ground and the first excited states show different behaviors. The ground state energy is a monotonically decreasing function of $d$ in the limit $d\rightarrow0$. The decrease of the ground state energy occurs because the impurity wave function can localize around both of the vortices and its length scale increases compared to the single vortex background. Namely, sharing the impurity between two vortices gives rises to an attractive force. This is nothing but a {\it covalent bonding}-like effect for the vortices. Since the $(1,0)$ wave function is axially symmetric, we may call this the $\sigma_{\text{s-s}}$ bond. Then, we find a numerical fit of the covalent energy of the ground state as a function of $d\, (\gtrsim 5\xi)$ as
\begin{eqnarray}
{\cal E}_{(1,0)+(1,0)}(d) \simeq {\cal E}_{1,0}^{(1)} 
-11.57 \frac{\hbar^2}{2MN_2\xi^2}\bigg(\frac{\xi}{d}\bigg)^{3.49},
\label{eq:numerical_fit}
\end{eqnarray}
On the other hand, the energy eigenvalues of the first excited state increase as $d$ is decreased. This is the so-called \textit{anti-bonding} effect common for states of odd parity. The presence of a linear node prevents the wave function from becoming shallow, instead the wave function becomes steeper in this region. Next, we consider the case of $2Mg_{\rm 12}\xi^2n_1N_1/\hbar^2=5$. As is shown in Fig.~\ref{fig:bs_single}, each background vortex can localize the impurity not only as the ground state (1s) but also several higher excited states (2p, 2s, 3d, 3p, 3s). In general, a wave function of an excited state has a larger length scale compared to the ground state wave function, so covalent bonding effects start to appear for vortices placed with a larger separation compared to states with lower energy. Furthermore, when a bound state has non-zero angular momentum $(m\neq 0)$, its wave function is not axially symmetric but has directivity. As we will see in subsection \ref{sec:bfam}, these properties of the higher excited states will give us additional contributions to the covalent bonding effect. 
\begin{figure*}[t]
\begin{center}
\includegraphics[width=0.9\textwidth]{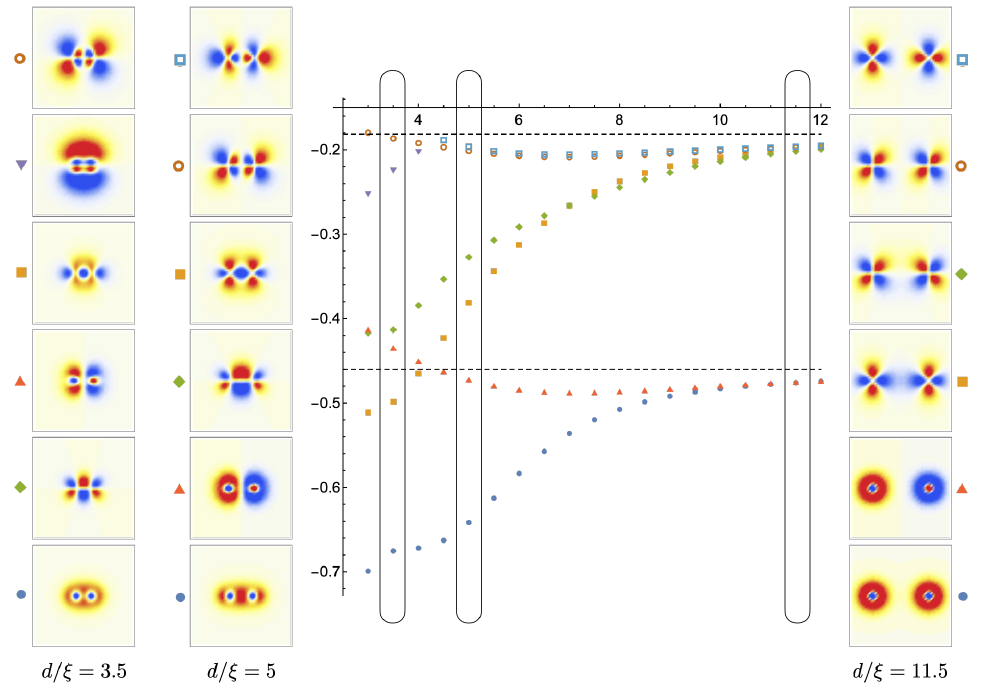}
\caption{(color online) The spectral flows of 2s-2s, and 3d-3d orbitals (the 3p-3p mixture partially appears) for $C_1=5$. The horizontal axis corresponds to the dimensionless distance $d/\xi$ while the vertical axis corresponds to the dimensionless energy eigenvalue $mN_2\xi^2{\cal E}/\hbar^2$. The lower dotted horizontal line shows ${mN_2\xi^2\cal E}_{\text{2s}}^{(1)}/\hbar^2\simeq-0.46$, and the upper one corresponds to $mN_2\xi^2{\cal E}_{\text{3d}}^{(1)}/\hbar^2\simeq-0.18$. The blue disks correspond to $|\mathlarger{\mathlarger{\circ}},+\rangle$ (Eq.~\eqref{eqn:2sp}), red triangles denote $|\triangle,-\rangle$ states (Eq.~\eqref{eqn:2sm}), orange squares denote $|\square,+\rangle$ (Eq.~\eqref{eqn:ppp}), green diamonds denote $|\mathlarger{\mathlarger{\diamond}},+\rangle$ (Eq.~\eqref{eqn:ttp}), brown circles denote $|\mathlarger{\mathlarger{\circ}},-\rangle$ (Eq.~\eqref{eqn:ttm}), blue squares denote $|\square,-\rangle$ (Eq.~\eqref{eqn:ppm}), while the purple upside-down triangles denote a $\delta_{p-p}$ state. In all cases, we show the wave functions for displacements $d/\xi=3.5,5,11.5$. The spatial region of the density plots is $x/\xi \in [-10,10]$ and $y/\xi \in [-10,10]$.}
\label{fig:bs_double_20_12}
\end{center}
\end{figure*}
Let us begin with the ground state and first excited state, namely the 1s-1s mixtures. The numerical results are shown in Fig.~\ref{fig:bs_double_10}. Qualitatively, there is no difference from the case with $2Mg_{\rm 12}\xi^2n_1N_1/\hbar^2=1$. The two states are almost degenerate in energy when the separation $d$ is sufficiently large. As $d$ decreases, the energy of both states decreases in a similar fashion from the asymptotic energy ${\cal E}_{1,0}^{(1)} \simeq -2.85$. Then, when $d$ is small enough, the spectrum undergoes splitting. The ground state energy continuously lowers due to the $\sigma_{s-s}$ bonding, while the energy of the first excited state stops lowering, and instead starts increasing due to the \textit{anti-bonding} effect. As discussed above, the wave functions for larger $g_{\rm 12}$ are spatially smaller than those for smaller $g_{\rm 12}$. Therefore, the spectral splitting occurs at relatively small $d\,(\lesssim 4.5\xi)$ in comparison with the case of $C_1=1$. 

Although we studied the impurity bound states and bonding of vortex-vortex states, the choice of the vortex charge $q$ will in fact not affect the bound-states, since the impurity feels the condensate density through a potential that depends on the density of the first component, (see Eq.~\eqref{eq:schrodinger}). However, the dynamics of the coupled system would be quite different depending on the choice of $q$.

\subsection{\label{sec:bfam}Bonding at finite angular momentum}
Let us next look at higher excited states (2p) in which two independent states $(n,m)=(1,\pm 1)$ can coexist. Fig.~\ref{fig:bs_double_11} shows the numerical results. When the vortex-vortex separation $d\gg\xi$, there are four degenerate states. Each state has directivity like a dipole, so that a total of four states can be geometrically constructed, (see right-most four panels of Fig.~\ref{fig:bs_double_11}). We denote the individual 2p wave functions by the states $|\text{2p},{\leftarrow}\rangle,|\text{2p},{\rightarrow}\rangle,|\text{2p},\uparrow\rangle,|\text{2p},\downarrow\rangle$ expressing the direction of individual dipole axes. We consider the following products (viz. Eq.~\eqref{eqn:vprod}) of these four states
\begin{subequations}
\begin{align}
\label{eqn:circ}|\mathlarger{\mathlarger{\circ}}\rangle&=|\text{2p},\leftarrow\rangle\otimes|\text{2p},\rightarrow\rangle,\\
\label{eqn:tri}|\triangle\rangle&=|\text{2p},\leftarrow\rangle\otimes|\text{2p},\leftarrow\rangle,\\
\label{eqn:sq}|\square\rangle&=|\text{2p},\uparrow\rangle\otimes|\text{2p},\uparrow\rangle,\\
\label{eqn:dia}|\mathlarger{\diamond}\rangle&=|\text{2p},\uparrow\rangle\otimes|\text{2p},\downarrow\rangle.
\end{align}
\end{subequations}
When the two vortices are well separated, the four states given by Eq.~\eqref{eqn:circ}-\eqref{eqn:dia} are almost degenerate. However, they are not exactly degenerate as long as the distance $d$ is finite. Then, ordering these four states by their energy, we have the lowest (second excited) state $|\mathlarger{\mathlarger{\circ}}\rangle$, Eq.~\eqref{eqn:circ} (here the arrows indicate the direction on the bond axis, a blue disk in Fig.~\ref{fig:bs_double_11}), the second from below (the third excited) state is denoted by $|\triangle\rangle$ (orange triangle, Eq.~\eqref{eqn:tri}), while the third from below (fourth excited state) state is denoted by $|\square\rangle$ where the arrows are orthogonal to the bond axis (yellow squares, Eq.~\eqref{eqn:sq}). The fifth excited state is denoted by $|\mathlarger{\diamond}\rangle$ where the arrows are orthogonal to the bond axis (green diamonds, Eq.~\eqref{eqn:dia}).

The eigenenergies of the four states simultaneously decrease as $d$ decreases with respect to their energy, with their order unchanged. However, on the interval $6 < d/\xi < 6.5$, level crossings occur. Namely, the state $|\square\rangle$ becomes the third from fourth while the state $|\triangle\rangle$ is lifted from the third to fourth state, (see inset of Fig.~\ref{fig:bs_double_11}). The latter state $|\triangle\rangle$ is further lifted and becomes the fifth excited state as $d$ decreases. The state $|\mathlarger{\mathlarger{\circ}}\rangle$ does not participate in mixing, remaining at the lowest level. The $|\mathlarger{\mathlarger{\circ}}\rangle$ states energy suddenly decreases, caused by the increasing overlap of the individual lobes of each vortex. We denote this $\sigma_{p-p}$ bonding. Following this, the remaining three states also split around $d/\xi \lesssim 5$. The energy of the second lowest state $|\square\rangle$ also continuously decreases since two lobes of one orbital can hybridize with two lobes of the other. We denote this $\pi_{p-p}$ bonding. On the contrary, overlaps of lobes cannot occur due to their geometry for the remaining two states $|\mathlarger{\mathlarger{\diamond}}\rangle$ and $|\triangle\rangle$. For these states smaller displacements $d$ cause their respective energies to increase due to the {\it anti-bonding} effect.
 
Finally, we explain the higher excited states involving the 2s and 3d orbitals, which are shown in Fig.~\ref{fig:bs_double_20_12}. The two states 2s are denoted by
\begin{align}
\label{eqn:2sp}|\mathlarger{\mathlarger{\circ}},+\rangle&=|\text{2s},+\rangle\otimes|\text{2s},+\rangle,\\
\label{eqn:2sm}|\triangle,-\rangle&=|+,\text{2s}\rangle\otimes|\text{2s},+\rangle,
\end{align}
and the interchanged labels appearing in the first ket on the r.h.s. of Eq.~\eqref{eqn:2sm} indicates the odd parity version of $|\text{2s},+\rangle$, a general notation we adopt for the remainder of this work. When the vortex displacement $d$ is large, the mixtures of the two 2s orbitals have almost degenerate energies which asymptotically converge to $2MN_2\xi^2{\cal E}_{\text{2s}}^{(1)}/\hbar^2 \simeq -0.46$. Since the 2s orbitals are axially symmetric, the spectral flows of these almost degenerate states are qualitatively similar to the 1s-1s bound states, as explained above. However, a difference does arise from the length scales of the wave functions. Since the 2s orbital is larger than the 1s one, the splitting of the two degenerate states due to the $\sigma_{s-s}$ anti-bonding begins at relatively large $d\, (\lesssim 9\xi)$ displacement for the 2s state. However, by carefully examining the spectral flow, we find that the $\sigma_{s-s}$ bonding occurs in two stages. The first is due to overlap of the wave functions outer shells $(d \lesssim 9\xi)$, while the second occurs instead due to the overlap of the wave functions inner shell $(d\lesssim 3.5\xi)$. On the other hand, two vortices with 3d orbitals have four almost degenerate states whose energies asymptotically converge to $mN_2\xi^2{\cal E}_{\text{3d}}^{(1)}/\hbar^2 \simeq -0.18$. Since the 3d orbital has four lobes, we symbolically express these states as $|\text{3d},+\rangle$ and $|\text{3d},\times\rangle$. Amongst these four 3d states, the lowest energy state is denoted 
\begin{equation}
\label{eqn:ppp}|\square,+\rangle=|\text{3d},+\rangle\otimes|\text{3d},+\rangle,
\end{equation}
where a single lobe of one orbital facing a single lobe of another has common color (an orange square in Fig.~\ref{fig:bs_double_20_12}). Then, the second lowest energy state is 
\begin{equation}
\label{eqn:ttp}|\mathlarger{\mathlarger{\diamond}},+\rangle=|\text{3d},\times\rangle\otimes|\text{3d},\times\rangle,
\end{equation}
where two lobes of one orbital are facing towards the two lobes of another, and have a common color (green diamond in Fig.~\ref{fig:bs_double_20_12}). The energy eigenvalues of both of these states decreases when the vortex separation $d/\xi$ approaches (but cannot reach) zero separation. On the other hand, the odd parity states
\begin{equation}
\label{eqn:ttm}|\mathlarger{\mathlarger{\circ}},-\rangle=|\times,\text{3d}\rangle\otimes|\text{3d},\times\rangle
\end{equation}
 correspond to brown circles, while 
 \begin{equation}
 \label{eqn:ppm}|\square,-\rangle=|+,\text{3d}\rangle\otimes|\text{3d},+\rangle,
 \end{equation}
correspond to open blue squares. 
The energy of the $|\mathlarger{\mathlarger{\circ}},-\rangle$ and $|\square,-\rangle$ states are lifted by the anti-bonding effect in the region where $d$ is small. In the asymptotic limit ($d\gg\xi$), the four states are almost, but not exactly degenerate. The observed ordering is ${\cal E}_{\square}^{+} < {\cal E}_{\diamond}^{+} < {\cal E}_{\circ}^{-} < {\cal E}_{\square}^{-}$. However, as $d$ decreases, level crossings occur between the $|\square,+\rangle$ and $|\mathlarger{\mathlarger{\diamond}},+\rangle$ states. In the region $d \lesssim 5\xi$, the lowest energy state is $|\mathlarger{\mathlarger{\diamond}},+\rangle$. The lowering of energy is due to the $\sigma_{p-p}$ bonding for $|\square,+\rangle$ while for the $|\mathlarger{\mathlarger{\diamond}},+\rangle$ this effect originates instead from $\pi_{\text{d-d}}$ bonding. Therefore, the level crossing implies that the latter bonding is more efficient than the former one. In the region where $d$ is small enough, these states are relatively smaller than the anti-$\sigma_{s-s}$ bonded 2s-2s mixture with odd parity. We also find another type of covalent bond, $\delta_{p-p}$ which is a bound state of 3p orbitals (upside-down purple triangle) appearing second from the top ($d \lesssim 4\xi$) in Fig.~\ref{fig:bs_double_20_12}. This involves bonding among four lobes, each from a 3p orbital. Additionally, there are further higher excited states of 3p-3p and 3s-3s which we do not show here.

\section{\label{sec:cc}Conclusions}

In this work, we have modelled the interaction of a superfluid vortex with quantum impurity atoms. This has been accomplished using a two component (binary) Gross-Pitaevskii formalism, where the superfluid-impurity interaction has been handled via a simple density-density interaction. The superfluid-impurity interaction has been characterized in the mass-imbalanced regime, and it was found that the presence of the impurity atoms can have a dramatic effect on the superfluid vortex; leading to distorted vortex profiles even at modest atom imbalances. The properties of vortex pairs have also been studied in the presence of impurities, and it was shown that there is an attractive splitting of the impurities energy that depends on the signs of the vortex pairs. The splitting of the impurities energy has been observed to be greater at smaller imbalances. We also investigated the time-dependent dynamics of the vortex-impurity system, observing that the energies of the impurities are not significantly altered by dynamics, as long as the separation of the vortices is several healing lengths.

In the second part of this work we considered the excited states of the coupled system. Using a Pad\'e approximation, a model for the impurity potential generated by the vortex was constructed that allowed us to diagonalize the impurity component in the limit of small atom number imbalances, subsequently allowing the examination of the excited states of the system. In general, the number of accessible excited states depends on the details of the coupling between the two components as well as the particular choice of atom number imbalance. For deeper potentials, more excited states become accessible. The impurity orbitals of this two-dimensional system possess a two-dimensional hydrogen-like character, with higher excited states carrying finite angular momentum, and being more diffuse in character as the zero (ionization)  energy limit is approached. The excited impurity orbitals were  then used to construct the eigenstates of a pair of like-sign vortices. The spectral flows of the impurities lowest energy excited states revealed an attractive power law behaviour, raising the possibility of covalent bonds between vortices. The spectral flows of the excited (angular momentum carrying) states also revealed a rich behaviour. Chemical bonding of pairs of vortices possessing one or two units of angular momentum display physical effects analogous to their chemical counterparts, such as the appearance of bonding and anti-bonding states, as well as spectral level crossings.

There are several interesting routes for future investigations. Extending this analysis to the case of fermions is a novel and worthwhile avenue, since this system is a good candidate for producing a novel matter-wave-trapped degenerate Fermi gas, analogously to the experiment of DeSalvo et al., Ref.~\cite{desalvo_2017}. The impurity states of this system also represent good candidates for quantum information applications, such as long-lived qubits analogously to those proposed for dark solitons \cite{shaukat_2017,shaukat_2018}. Our analysis of the impurity mediated chemical bonds could also be used as a basis to construct more elaborate forms of controllable synthetic matter, such as Toda lattices \cite{toda_1967}, vortex-impurity crystals and the impurities associated superfluid-insulator transitions. Further, collisions between multiple vortex carrying impurities would allow for the simulation of synthetic chemical reactions. Our work also represents a new tool in the growing field of atomtronics \cite{amico_2020}, where the ability to construct synthetic chemical bonds yields a useful new paradigm in this emerging discipline.   

\section*{Acknowledgements}
This work is supported in part by Japan Society of Promotion of Science (JSPS) Grant-in-Aid for Scientific Research (KAKENHI Grants No. JP20K14376 (M.~Edmonds), No. JP19K03839 (M.~Eto), and JP18H01217 (M.~N.)), and also by MEXT KAKENHI Grant-in-Aid for Scientific Research on Innovative Areas, Discrete Geometric Analysis for Materials Design, No. JP17H06462 (M.~Eto) from the MEXT of Japan.
\appendix
\section{\label{app:ints}Variational integrals}
In this appendix we list the integrals required to evaluate the various contributions to the variational energy, Eqs.~\eqref{eqn:ven1}-\eqref{eqn:ven4} in Sec.~\ref{sec:vc} of the main text. Below we explain the origin of the individual integrals  
\begin{widetext}
\begin{align}
\label{eqn:app1}&\int^{\infty}_{0}du u\bigg[\frac{1}{\beta^2(\beta+1)^2} - \bigg(\frac{1}{\beta+\exp(-u^2)} - \frac{1}{\beta+1}\bigg)^2\bigg]=\frac{(\beta-1)\ln(\frac{\beta}{\beta+1})+1}{2\beta^2(\beta+1)},\\
\label{eqn:app2}&\int^{\infty}_{0}du u\bigg[\frac{\partial}{\partial u}\frac{1}{\beta+\exp(-u^2)}\bigg]^2=\frac{1}{3\beta^2}\bigg[\ln\bigg(\frac{\beta+1}{\beta}\bigg)-\frac{\beta}{(\beta+1)^2}\bigg],\\
\label{eqn:app3}&\int^{\infty}_{0} du u\bigg[\frac{1}{\beta^2(\beta+1)^2}-\bigg(\frac{1}{\beta+\exp(-u^2)}-\frac{1}{\beta+1}\bigg)^2\bigg]^2=\frac{6(\beta+1)(\beta-1)^2\ln(\frac{\beta+1}{\beta})+1+3\beta(3-2\beta)}{12\beta^4(\beta+1)^3}.
\end{align}
\end{widetext}
To calculate the normalization of the variational function $\psi_{\rm 1}^{\rm var}(x,y)$, the first integral Eq.~\eqref{eqn:app1} is used. To evaluate the contribution from the density of the variational kinetic energy Eq.~\eqref{eqn:ekin} for the vortex, the second integral given by Eq.~\eqref{eqn:app2} is required. Finally, Eq.~\eqref{eqn:app3} is needed to calculate the van der Waals contribution arising from Eq.~\eqref{eqn:vdw1}.  

\end{document}